\documentclass[11pt,a4paper]{article}  	
\usepackage{geometry}                		

\usepackage{amssymb,amsmath}
\usepackage{float}
\usepackage{tabularx}
\usepackage{graphicx}
\usepackage{t1enc}
\usepackage[icelandic,english]{babel}
\usepackage{graphicx}
\usepackage{amsmath}
\usepackage{amsthm}
\usepackage{amssymb}
\usepackage{multirow}
\usepackage{enumitem}
\usepackage{t1enc}
\usepackage{amssymb}
\usepackage{color}
\usepackage{latexsym}
\usepackage{epic}
\usepackage{epsfig}
\usepackage{eepic}
\usepackage{amsmath}
\usepackage{enumerate}
\usepackage{graphicx}
\usepackage{multicol}	
\usepackage{natbib}
\usepackage{tabularx}
\usepackage{graphicx}
\usepackage{caption}
\usepackage{subcaption}
\numberwithin{equation}{section}

\newcommand{\Matrix}[1]{\begin{pmatrix} #1 \end{pmatrix}}

\newcommand{\fat}[1]{\boldsymbol{#1}}

\newcommand{\inv}{^{-1}}
\newcommand{\ndist}[2]{\mathcal{N}\left(#1,#2\right)} 
\newcommand{\trp}{^\mathsf{T}}

\newcommand{\ei}{\end{itemize}}

\usepackage{subcaption}
\usepackage{algorithm}
\usepackage{algorithmic}

\usepackage[toc,page]{appendix}	

\newtheorem{theorem}{Theorem}
\newtheorem{lemma}[theorem]{Lemma}

\newtheorem{corollary}[theorem]{Corollary}

\author{  \'Oli P\'all Geirsson, Birgir Hrafnkelsson and Helgi Sigurðarson \\
      Department of Mathematics\\
      Faculty of Physical Sciences\\
    School of Engineering and Natural Sciences\\
    University of Iceland \\
     and \\
    Daniel Simpson \\
    Department of Statistics\\
    University of Warwick, Coventry}
\title{The MCMC split sampler: A block Gibbs sampling scheme for latent Gaussian models}

\begin{document}
\maketitle

\begin{abstract}
A novel computationally efficient Markov chain Monte Carlo (MCMC) scheme for latent Gaussian models (LGMs) is proposed in this paper. The sampling scheme is a two block Gibbs sampling scheme designed to exploit the model structure of LGMs. We refer to the proposed sampling scheme as the MCMC split sampler. The principle idea behind the MCMC split sampler is to split the latent Gaussian parameters into two vectors. The former vector consists of latent parameters which appear in the data density function, while the latter vector consists of latent parameters which do not appear in it. The former vector is placed in the first block of the proposed sampling scheme and the latter vector is placed in the second block along with any potential hyperparameters. The resulting conditional posterior density functions within the blocks allow the MCMC split sampler to handle, by design,  LGMs with latent models imposed on more than just the mean structure of the data density function. The MCMC split sampler is also designed to be applicable for any choice of a parametric data density function. Moreover, it scales well in terms of computational efficiency when the dimension of the latent model increase. 

\end{abstract}

\section{Introduction}
\label{intro}
Latent Gaussian models (LGMs) form a flexible subclass of Bayesian hierarchical models and have become popular in many areas of statistics and various fields of applications, as LGMs are practical from a statistical modeling point of view and readily interpretable. For example, LGMs play an important role in spatial statistics, see \cite{cressie1993statistics, diggle1998model, delfiner2009geostatistics}; statistical climatology \citep{cooley2007bayesian,guttorp2006studies}; disease mapping \citep{pettitt2002conditional, lawson2013bayesian}; stochastic volatility models \citep{martino2011estimating}; and hydrology \citep{schaefli2007quantifying}, to name a few. Moreover, LGMs can be viewed as a specific extension of structured additive regression models \citep{fahrmeir1994multivariate, rue2009approximate}, in the sense that, the data density function of each data point can depend on more than a single linear functional of the latent field through more than just the mean structure, as discussed in \cite{martins2013discussion}.

Although LGMs are well suited from a statistical modeling point of view their posterior inference becomes computationally challenging when latent models are desired for more than just the mean structure of the data density function \citep{martins2013discussion}, when the number of parameters associated with the latent model increase; or when the data density function is non-Gaussian. The aim of this paper is to propose a novel computationally efficient Markov chain Monte Carlo (MCMC) scheme which serves to address these computational issues. The proposed sampling scheme is referred to as the \emph{MCMC split sampler} in this paper. It is designed to handle LGMs where latent models are imposed on more than just the mean structure of the likelihood. It scales well in terms of computational efficiency when the dimensions of the latent models increase and it is applicable for any choice of a parametric data density function. The main novelty of the MCMC split sampler lies in how the model parameters of a LGM are split into two blocks. As a result of the proposed blocking scheme, one of the blocks exploits the latent Gaussian structure in a natural way and becomes invariant of the data density function. 

Markov chain Monte Carlo (MCMC) methods form the backbone of modern Bayesian posterior inference and are, in principle, applicable to almost any Bayesian model.  However, the mixing and convergence properties of the MCMC chains can be poor for involved models structures and large data sets if parameters that are dependent in the posterior are not dealt with properly, see for example \cite{murray2010slice}. In particular, the mixing and convergence properties of the popular single site updating strategy can be extremely poor due to strong dependencies of parameters in the posterior distribution as discussed in \cite{knorr2002block}. 
Several MCMC sampling strategies have been suggested for Bayesian hierarchical models to improve the mixing properties of MCMC algorithms. For example, methods based on approximate diffusions such as the Metropolis Adjusted Langevin algorithm (MALA), see \cite{roberts1998optimal}; methods based on Hamiltonian mechanics (HMC), suggested by \cite{neal1993probabilistic}, which use the gradient of the target density to drive the proposal mechanism toward regions of higher posterior density; manifold methods, proposed by \cite{girolami2011riemann}, which provide a systematic way of designing proposal densities for MALA and HMC by making use of the gradient and curvature information of the target density; and various block sampling strategies such as the one block updating strategy of \citet{knorr2002block}. \citet{filippone2013comparative} conducted a detailed comparison of these methods for LGMs and found that the single block strategy of  \citet{knorr2002block},  in which the latent field and its corresponding hyperparameters are updated jointly in a single block,  performed best in most situations.  Furthermore, by using numerical methods for fast sampling of Gaussian Markov random fields (GMRFs) \citep{rue2001fast}, the single block sampler can be implemented with a low computational cost for LGMs. However, using only a single block sampler for LGMs with a non-Gaussian likelihood and high dimensional latent fields can be problematic, as parameters accepted in regions of low posterior probability can cause the  MCMC chain to get stuck.

Alternative to MCMC methods are deterministic approximate posterior inference methods, such as the Integrated nested Laplace approximation (INLA) \citep{rue2009approximate}. INLA is a fast approximate inference method for LGMs in which the data density of each data point only depends on a single linear functional of the latent field.  While this assumption holds in many practical cases, there are many models in which we want the latent field to enter the data density of a single observation through two or more parameters, see \cite{kneib2013beyond} and \cite{martins2013discussion} for further discussion. For example in \cite{hrafnkelsson2012spatial} and \cite{geirsson2015computationally}, latent Gaussian spatial models were imposed on the location, scale and shape structure of the data density function.

The MCMC split sampler is a two block Gibbs sampling scheme \citep{geman1984stochastic, casella1992explaining} designed for LGMs, which addresses the aforementioned inference problems. The MCMC split sampler is based on the following model setup for LGMs, to which we adhere to in this paper. 
   \begin{description}
 \item[Data-level:] The observations $\fat{y}$ depend on the latent field $\fat{x}$, through some choice of data distribution with a data density function  $\pi (\fat y \mid \fat x)$.
 \item[Latent level:] The prior for the latent field $\fat{x}$ is Gaussian and is potentially dependent on \mbox{hyperparameters $\fat \theta$,} with a density function
     \begin{displaymath}
         \pi(\fat{x}\mid \fat{\theta}) \;= {\mathcal N}\left(\fat x \mid 
         \fat{\mu}(\fat{\theta}),
         \fat{Q}(\fat{\theta})^{-1}\right).
     \end{displaymath}
 \item[Hyperparameter level:] A prior distribution is assigned for the hyperparameters $\fat \theta$, with a density function $\pi (\fat \theta)$.
 \end{description}
The principle idea behind the MCMC split sampler is to split the latent Gaussian parameters $\fat x$ into two vectors, $\fat \eta$ and $\fat \nu$, where $\fat \eta$ consists of elements that appear in the data density function and $\fat \nu$ consists of elements that do not appear in it. Thus, the data $\fat y$ become conditionally independent of $(\fat \nu, \fat \theta)$ conditioned on $\fat \eta$, that is, $\pi( \fat y \mid \fat x, \fat \theta) = \pi (\fat y \mid \fat \eta)$. For the posterior inference, all the model parameters are grouped into two blocks. That is, $\fat \eta$ is placed in a block we refer to as the \emph{data-rich} block in this paper, while both $\fat \nu$ and the hyperparameters $\fat \theta$ are placed in another block referred to as the \emph{data-poor} block. A Gibbs sampling strategy is then implemented for each block, conditioned on the other block.

In many practical applications with non-Gaussian data density functions, especially in the field of spatial statistics, the vector  $\fat \eta$ in the data-rich block has a complicated but low-dimensional conditional posterior structure, while the parameter vector $\fat \nu$ in the data-poor block is often of much higher dimension than that of the parameters in the data-poor block \citep{hrafnkelsson2012spatial}. Therefore, by using the proposed blocking scheme the potentially computationally demanding conditional posterior density in the data-rich block contains a minimum number of necessary parameters. Furthermore, the conditional posterior density of $\fat \nu$ becomes conditionally Gaussian conditioned on the parameter vector $\fat \eta$ and the hyperparameters $\fat \theta$. This designed structure, that is, the minimal dimension of the parameters in the data-rich block and the conditional Gaussian  posterior structure within the data-poor block, is exploited to implement computationally efficient sampling schemes within the blocks. Further, the proposed scheme scales well when the dimension of $\fat \nu$ increases, as discussed in the ensuing paragraph.

%

The MCMC split sampler is modular by design such that, in principle, any efficient MCMC sampler can be implemented for each block. In this paper, we propose computationally efficient sampling strategies that are tailored to the particular conditional model structure of each block. Within the data-rich block we present a strategy based on the gradient and curvature information of the target density that results in an independence proposal mechanism as discussed in \cite{rue2005gaussian}. Conditional independence resulting from the model structure within the block can be utilized in some cases to increase acceptance in the Metropolis-Hastings algorithm \citep{metropolis1953equation,hastings1970monte}. In order to update the data-poor block, a modified version of the fast single block updater of \citet{knorr2002block} is proposed, which exploits the fact that the conditional posterior of the vector $\fat \nu$ is Gaussian conditioned on $\fat \eta$ and $\fat \theta$. This step is invariant of the choice of a data density function as the the data-poor block is updated conditioned on the data-rich block. Moreover, if the latent field $\fat x$ is a GMRF with a sparse precision structure, the sampling strategy for the data-poor block is shown to conserve the sparse GMRF precision structure, allowing for fast sampling of the corresponding GMRF.

The paper is organized as follows. Section 2.1 and Section 2.2 are devoted to the motivation, introduction and the setup of the MCMC split sampler. The proposed sampling schemes within the data-rich and data-poor blocks are presented in Sections 2.3 and Section 2.4, respectively. Examples on the implementation of the MCMC split sampler are given in Section 3. In Section 3.1 we present a LGM with a latent spatial model structure on mean and log-variance parameters, and show how the MCMC split sampler scales well as the dimensions of the latent parameters in the data-poor block increase. An example on extremes based on a simulated data set is given in Section 3.2, where the focus is on a LGM where latent models are imposed on all three parameters of the generalized extreme value distribution. In Appendix A, we give an extension to the sampling scheme proposed in Section 2.3, which is applicable  if conditional independence assumptions are imposed on the data data density function. Lastly, in Appendix B, we show the necessary proofs for the main results in the paper.

\section{The MCMC split sampler }
\label{MCMCsplit}

\subsection{Motivation and model setup}

Consider, as a motivation and without loss of generality, a data density function $\pi(\fat y \mid \fat \mu, \fat \tau)$ where $\fat \mu$ and $\fat \tau$ are vectors of location and log-scale parameters, respectively, which are modeled with latent Gaussian fields. That is, assume the following additive model structure 
\begin{align*}
	\fat \mu = \fat X_\mu \fat \beta_\mu + \fat A_\mu \fat u_\mu \text{,} \quad 
	\fat \tau = \fat X_\tau \fat \beta_\tau + \fat A_\tau \fat u_\tau 
\end{align*}
where $\fat X_\mu$ and $\fat X_\tau$ are fixed design matrices; $\fat \beta_\mu$ and $\fat \beta_\tau$ are the corresponding weights;  $\fat A_\mu$ and $\fat A_\tau$ are fixed matrices; and $\fat u_\mu$ and $\fat u_\tau$ are structured random effects. In order to increase computational stability in the posterior inference calculations and for mathematical derivations of the MCMC split sampler, we introduce unstructured random effects, $\fat \epsilon_\mu$ and $\fat \epsilon_\tau$, to the model. That is,
\begin{align}
\label{dform}
	\fat \mu = \fat X_\mu \fat \beta_\mu + \fat A_\mu \fat u_\mu + \fat \epsilon_\mu  \text{,} \quad 
	\fat \tau = \fat X_\tau \fat \beta_\tau + \fat A_\tau \fat u_\tau + \fat \epsilon_\tau.
\end{align}
Small variances can be imposed \emph{a priori} on the unstructured random effects $\fat \epsilon_\mu$ and $\fat \epsilon_\tau$  if they are not desired in the model. However, adding the unstructured random effects is reasonable in many cases from a statistical modeling point of view as they serve as error terms for the latent models. Furthermore, adding the $\fat \epsilon_\mu$ and $\fat \epsilon_\tau$ terms to the latent models yields an analogous latent model structure as implied by the structured additive regression in \cite{fahrmeir1994multivariate} where only the mean is linked to a structured additive predictor through a link function.

Assign the following Gaussian prior density functions to the latent model parameters 

\begin{align}
\label{ddpp}
  \pi(\fat \beta_\mu) &= \mathcal{N}(\fat \beta_\mu\mid\fat \mu_{\mu \beta},  \fat Q_{\mu \beta}^{-1}), \quad  \pi(\fat \beta_\tau) = \mathcal{N}(\fat \beta_\tau\mid\fat \mu_{\tau \beta},  \fat Q_{\tau \beta}^{-1}) \nonumber \\ 
  \pi(\fat u_\mu) &= \mathcal{N}(\fat u_\mu\mid\fat \mu_{\mu u} , \fat Q_{\mu u}^{-1}), \quad \pi(\fat u_\tau) = \mathcal{N}(\fat u_\tau\mid\fat \mu_{\tau u} , \fat Q_{\tau u}^{-1}) \nonumber\\
   \pi(\fat \epsilon_\mu) &=\mathcal{N}(\fat \epsilon_\mu\mid \fat 0, \fat Q_{\mu \epsilon}^{-1}), \quad \quad \text{ } \pi(\fat \epsilon_\tau) =\mathcal{N}(\fat \epsilon_\tau\mid \fat 0, \fat Q_{\tau \epsilon}^{-1}) 
\end{align}
where parameters of the prior density functions can potentially depend of a set of hyperparameters $\fat \theta$, and $\fat Q_{\mu \epsilon}^{-1}$ and $\fat Q_{\tau \epsilon}^{-1}$ are diagonal matrices. 

As the vector $\fat \mu$ in equation (\ref{dform}) is a linear combination of $\fat \beta_\mu$, $\fat u_\mu$ and $\fat \epsilon_\mu$, it is equivalent to obtain MCMC samples from the posterior distribution of $(\fat \mu, \fat \beta_\mu, \fat u_\mu)$ and from the posterior distribution of $(\fat \beta_\mu, \fat u_\mu, \fat \epsilon_\mu)$. Analogous argument holds for the log-scale parameters. The MCMC split sampler is designed to obtain MCMC samples from the posterior distribution of $(\fat \mu, \fat \tau, \fat \beta_\mu, \fat \beta_\tau, \fat u_\mu, \fat u_\tau)$ as opposed to $(\fat \beta_\mu, \fat \beta_\tau, \fat u_\mu, \fat u_\tau,\fat \epsilon_\mu,\fat \epsilon_\tau)$ as in the former parameterization only the vector $(\fat \mu, \fat \tau)$ enters the data density function, while all the elements of the latter vector enter the data density function in the latter parameterization. This parameterization for posterior inference is along the lines of the posterior inference scheme proposed in \cite{rue2009approximate}. Thus, define
\[
	\fat \eta = \left( \fat \mu, \fat \tau \right)\trp, \quad \fat \nu = \left( \fat \beta_\mu, \fat u_\mu, \fat \beta_\tau, \fat u_\tau\right)\trp
\]
which will act as the splitting of the parameters of the latent field. 

The latent model structure in (\ref{dform}) and the prior distributions in (\ref{ddpp}) can be written in a joint matrix form, which forms the basis for the derivation of the MCMC split sampler. Define the following matrices and vectors
\begin{align*}
  \fat Z &= \begin{pmatrix} 
    \fat X_\mu & \fat A_\mu & \cdot & \cdot \\
    \cdot & \cdot & \fat X_\tau & \fat A_\tau \\
  \end{pmatrix}
  \text{, }
  \quad
  \fat \epsilon  = \begin{pmatrix} 
    \fat \epsilon_\mu  \\
    \fat \epsilon_\tau  \\
    \end{pmatrix}
      \text{, }
 \quad
  \fat Q_\epsilon = \begin{pmatrix} 
    \fat Q_{\mu\epsilon}  & \cdot  \\
    \cdot & \fat Q_{\tau \epsilon}   \\
    \end{pmatrix}
\end{align*}
and group the following parameters and matrices together 
\[
  \fat \mu_\nu = \begin{pmatrix} 
    \fat \mu_{\mu \beta }\\
    \fat \mu_{\mu u }\\
     \fat \mu_{\tau \beta }\\
      \fat \mu_{\tau u }\\
  \end{pmatrix}
  \text{, }
  \quad
  \fat Q_\nu = \begin{pmatrix} 
    \fat Q_{\mu\beta}  & \cdot & \cdot & \cdot \\
    \cdot  & \fat Q_{\mu u} & \cdot & \cdot \\
    \cdot  & \cdot &  \fat Q_{\tau\beta}  & \cdot \\
    \cdot  & \cdot & \cdot  &  \fat Q_{\tau u} \\
    \end{pmatrix}
\]
where the dotted entries denote zero entries. The additive model structure implied by (\ref{dform}) for both latent parameters is thus equivalent to the matrix form
\begin{align}
\label{mform}
  \fat \eta = \fat Z \fat \nu + \fat \epsilon 
\end{align}
and the Gaussian prior assumptions in (\ref{ddpp}) are equivalent to
\begin{align}
\label{priors}
\pi\left(\fat{\eta}\mid\fat \nu\right) &= \ndist{\fat{\eta}\mid \fat Z \fat\nu}{\fat{Q}_\epsilon\inv}, \\
 \pi\left(\fat \nu\right) &= \ndist{\fat \nu \mid \fat \mu_\nu}{\fat{Q}_\nu\inv}. \notag
\end{align}

As the data density function and the corresponding parameters were arbitrarily chosen above, analogous derivations can be carried out for any parametric data density function and any of its parameters. For example, in addition to imposing latent Gaussian models on the location and log-scale parameters of the generalized extreme value distribution a latent Gaussian model can also be imposed on the shape parameter, see Section \ref{Flood} for details.  Therefore, equations (\ref{mform}) and (\ref{priors}) are general in the sense that most of LGMs used in practice can be expressed in the same form. We will thus adapt equations (\ref{mform}) and (\ref{priors}) as a general setup for the latent model structures for LGMs henceforth in this paper. 
 
The following lemma, based on known results, plays a vital role in the implementation of the MCMC split sampler. 

\begin{lemma}
\label{setnA}
Assume the distribution assumptions given in (\ref{priors}), for any mean vector $\fat \mu_{\nu}$, fixed matrix $\fat Z$, and precision matrices $\fat Q_{\epsilon}$ and $\fat Q_{\nu}$. The joint prior density function of $(\fat \eta, \fat \nu)$ is then Gaussian of the form
\begin{align}\label{joint3}
  \pi{\Matrix{\fat \eta \\ \fat \nu}} = \ndist{{\Matrix{\fat \eta \\ \fat \nu}} \bigg|  \Matrix{\fat Z\fat{\mu}_\nu\\ \fat{\mu}_\nu}}{\Matrix{\fat{Q}_\epsilon & -\fat{Q}_\epsilon\fat Z \\ -\fat Z\trp\fat{Q}_\epsilon & \fat{Q}_\nu+\fat Z\trp\fat{Q}_\epsilon\fat Z}\inv}
\end{align}
and the conditional density function of $\fat \nu$ conditioned on $\fat \eta$ becomes
\begin{align}
\label{condilater}
  \pi(\fat \nu \mid \fat \eta)  =  \mathcal N  \left( \fat\nu \Big | \fat Q_{\nu|\eta} ^{-1}(\fat Q_\nu \fat \mu_\nu + \fat Z\trp \fat Q_\epsilon \fat\eta) , \fat Q_{\nu|\eta} ^{-1}\right)
\end{align}
where $\fat Q_{\nu|\eta} = \fat{Q}_\nu+\fat Z\trp\fat{Q}_\epsilon\fat Z$.
\end{lemma}
See Appendix \ref{Proof_Gauss} for proof. Note that, as the vector $(\fat \eta, \fat \nu)$ is jointly Gaussian it can be viewed as the latent Gaussian vector $\fat x$ in the LGM setup.

\subsection{The sampling scheme} \label{sec:simple_sampling}
The vector $\fat \eta$ in (\ref{mform}) consists of the parameters of the latent field that explicitly enter the likelihood function while the vector $\fat \nu$ consists of the parameters of the latent field which do not enter it.  Therefore, the data vector $\fat y$ is conditionally independent of $\fat \nu$ conditioned on $\fat \eta$, that is $\pi( \fat y \mid \fat \eta, \fat \nu) = \pi (\fat y \mid \fat \eta)$. The parameters $\fat \eta$ and $\fat \nu$ are referred to as the data-rich and data-poor  components of the latent field, respectively, in this paper. The corresponding posterior distribution, where the data-poor components of the latent field are potentially dependent on a vector of hyperparameters $\fat \theta$, is thus proportional to
\begin{align}
\label{post_form}
   \pi (\fat \eta,\fat \nu, \fat \theta \mid \fat y ) \propto \pi (\fat y \mid \fat \eta) \pi (\fat \eta, \fat \nu \mid \fat \theta )\pi(\fat \theta).
\end{align}
Using the relationship in (\ref{post_form}), we propose the following  two block MCMC sampling scheme to obtain MCMC samples from the posterior density $\pi (\fat \eta,\fat \nu, \fat \theta \mid \fat y)$. The vector $\fat \eta$ is placed in the data-rich block, and the vectors $\fat \nu$ and  $\fat \theta$ are grouped together in the data-poor block. The MCMC split sampler obtains a sample from the posterior density $\pi (\fat \eta,\fat \nu, \fat \theta \mid \fat y)$ by sampling from one of the blocks conditioned on the other in a Gibbs sampling setting. That is, the $(k+1)$-th MCMC sample from the posterior density  $\pi(\fat \eta, \fat \nu, \fat \theta \mid \fat y)$ is obtained by using the following two block Gibbs sampling scheme 
\begin{description}
\item[Data-rich block:] sample $\fat \eta^{k+1}$ from 
$
\pi( \fat \eta \mid \fat y, \fat \nu^k, \fat \theta^k)
$
\item[Data-poor block:]  sample $ (\fat \nu^{k+1}, \fat \theta^{k+1})$ jointly from 
$
\pi( \fat \nu, \fat \theta\mid \fat y, \fat \eta^{k+1})
$
\end{description}
This scheme forms the basis of the MCMC split sampler. The potentially involved but often low-dimensional structure of the data-rich block is separated from the parameters in the data-poor block. By separating the two blocks, MCMC sampling strategies which exploit the conditional model structures can be implemented within each block in order to increase computational efficiency. Although any computationally efficient MCMC samplers are applicable within the blocks, 
we propose the following sampling schemes which are tailored for the conditional model structures of the blocks. The details of the proposed samplers for each block are summarized in Section \ref{sec:datarich} and Section \ref{sec:datapoor}.

\subsection{Sampler for the data-rich block}
\label{sec:datarich}
The conditional posterior density function $\pi(\fat \eta \mid \fat y, \fat \nu, \fat \theta)$ in the data-rich block is intractable in most applications. In order to obtain MCMC samples from the conditional posterior density function we propose the following Metropolis--Hasting type MCMC algorithm with a tailored independence proposal density \citep{rue2005gaussian}.  

To construct a computationally efficient independence proposal density, we approximate the conditional posterior density $\pi(\fat \eta \mid \fat y, \fat \nu, \fat \theta)$ with a Gaussian approximation evaluated at the mode of conditional posterior density. 
Using the logarithm of the conditional posterior, that is
 \begin{align}\label{logdens1}
  \log \pi ( \fat \eta \mid \fat y, \fat \nu, \fat \theta) =f( \fat \eta) -\frac{1}{2} \fat \eta\trp \fat Q_{\epsilon} \fat \eta +(\fat Q_{\epsilon} \fat Z\fat \nu)\trp \fat \eta  + \text{const}
\end{align}
where $f( \fat \eta) = \log \pi (\fat y\mid \fat \eta)$ for notational convenience, the following can be shown.
\begin{theorem}
\label{napprox}
The Gaussian approximation of the conditional posterior density \mbox{$\pi(\fat \eta \mid \fat y, \fat \nu,  \fat \theta)$} is given by
\begin{align}
\label{GaussApprox}
\tilde{\pi}(\fat \eta \mid \fat y, \fat \nu, \fat \theta) &= \mathcal{N} \left(\fat \eta \mid \fat\eta^0, (\fat Q_{\epsilon} -\fat H)\inv \right)
\end{align}
where $\fat \eta^0$ is the mode of the conditional posterior density $\pi ( \fat \eta \mid \fat y, \fat \nu, \fat \theta)$ and $\fat H$ is the the Hessian of the logarithm of conditional posterior evaluated at the mode, $\fat H = \nabla^2 f( \fat \eta^0 )$.
\end{theorem}
See Appendix \ref{Proof_Gauss} for proof. Note that, adding the additive unstructured error term $\fat \epsilon$ to the model in (\ref{ddpp}) prevents the the precision matrix in (\ref{GaussApprox}) from being singular and thus ensures numerical stability. 

As the Gaussian approximation in (\ref{GaussApprox}) is constructed at the conditional posterior mode $\fat\eta^0$, a proposal density $q$ for $\fat \eta$ based on (\ref{GaussApprox}) thus becomes invariant of the current position of $\fat \eta $ in the MCMC iteration. Therefore, the proposal density $q$ is an independence proposal density \citep{chib1995understanding,rue2005gaussian}. That is, in the $(k+1)$-th iteration the proposal density is invariant of $\fat\eta^k$, that is $q(\fat \eta^* \mid \fat\eta^k) = q(\fat\eta^*)$. 

When a new $\fat\eta^*$ is proposed with the independence proposal density in (\ref{GaussApprox}) in the $(k+1)$-th iteration, it is accepted with probability
\begin{align}\label{ra1}
  \alpha = \min\left\{1,\frac{ \pi ( \fat \eta^* \mid \fat y, \fat \nu, \fat \theta)}{ \pi ( \fat \eta^k \mid \fat y, \fat \nu, \fat \theta) }\cdot
  \frac{q(\fat \eta ^ k)} {q(\fat \eta ^ *) }  \right\}.
\end{align} 

The logarithm of the ratio in (\ref{ra1}) can be simplified, as stated in Lemma \ref{lemmaratio1}, in order to reduce computational cost. 
\begin{lemma}
\label{lemmaratio1}
Assume the proposal density $q$ implied by the Gaussian approximation in (\ref{GaussApprox}) for the data-rich block. The logarithm of the acceptance ratio given in (\ref{ra1}) can be simplified to
\begin{align}
\label{r1}
	r = &f( \fat \eta^*) -  \left( \frac1 2  (\fat \eta^*)\trp \fat H \fat   + \fat b\trp  \right) \fat \eta^*  -f( \fat \eta^k) +  \left( \frac1 2  (\fat \eta^k)\trp \fat H \fat   + \fat b\trp  \right) \fat \eta^k 
\end{align}
where $\fat b = \nabla f(\fat \eta^0) - \fat H \fat \eta^0$.

\end{lemma}
See Appendix \ref{Proof_lemma_datarich} for proof. As the gradient $\nabla f(\fat \eta^0)$ and Hessian $\fat H$ have already been calculated to obtain (\ref{GaussApprox}), the expression in (\ref{r1}) is computationally efficient to calculate. 

In many applications conditional independence assumptions are imposed on the data density function. That is, there exists a partition of $\fat \eta$ into subvectors $\fat \eta_i$,  such that
$
	\pi(\fat y \mid \fat \eta) = \prod_i \pi_i(\fat y_i \mid \fat \eta_i),
$
which is turn implies $f(\fat \eta) = \sum_i f_i(\fat \eta_{i} )$, where $f_i$ is the logarithm of the marginal data density function in the $i$-th partition. In some cases, a proposal density based on the Gaussian approximation in (\ref{GaussApprox}) can be a poor approximation of the conditional posterior density in some partition of $\fat \eta$. Updating the whole vector $\fat \eta$ in one block may then result in the MCMC chain getting stuck. As a result the computational efficiency of the sampler is reduced. In order to circumvent this issue and to retain the computational speed gained by using the Gaussian approximation in (\ref{GaussApprox}) as a proposal density, a modification can be made to the sampling scheme which utilizes the conditional independence of the partitions within the data-rich block. The details on the modification can be seen in Appendix \ref{CondInd}.  The resulting sampling scheme is outlined in Algorithm 1. Note that by choosing $I=1$ in Algorithm 1, the above sampling scheme without the conditional independence assumptions on the likelihood is obtained, while selecting $I\geq 2$ in Algorithm 1 assumes the aforementioned partitioning of $\fat \eta$ and that each $\fat \eta_i$ is accepted or rejected separately.

\begin{algorithm}[htp]
\label{alg_eta2}
\begin{algorithmic}[1]
\REQUIRE $(\fat \eta^k, \fat \nu^k)$
\STATE Find the mode $\fat \eta^0 = \underset{\fat \eta} {\operatorname*{arg\,max}} ~ \log \pi ( \fat \eta | \fat y,\fat \nu^k, \fat \theta) $
\STATE Calculate $\fat H = \nabla^2 f(\fat  \eta^0)$  and  $\fat b = \nabla f(\fat \eta^0) - \fat H \fat \eta^0$
\STATE Sample $\fat \eta^* \sim \mathcal{N} \left( \fat \eta^0, \left( \fat Q_\epsilon -\fat H \right)\inv \right)$ 
\STATE Calculate  $\fat \rho (\fat \eta^k)$ and $\fat \rho (\fat \eta^*)$, where $$\fat \rho(\fat \eta) = \left(-\frac1 2 \fat \eta\trp \fat H   - \fat b\trp \right) \circ \fat \eta$$ and $\circ$ denotes an entrywise product  
\FOR{$i=1,\ldots, I$}
\STATE Calculate $r_i =f_i( \fat \eta_i^*) + \fat \rho (\fat \eta^*)_i\trp \fat 1 - \left(f_i( \fat \eta_i^k) + \fat \rho (\fat \eta^k)_i\trp \fat 1 \right)$ 
\STATE Calculate $\alpha_i = \min\left\{ 1, \exp r_i \right\}$
\STATE Sample $u_i \sim \mathcal{U}(0,1)$
\IF{$\alpha_i > u_i$}
    \STATE $\fat \eta^{k+1}_i = \fat \eta^{*}_i $
  \ELSIF{$\alpha_i < u_i$}
  \STATE $\fat \eta^{k+1}_i = \fat \eta^{k}_i $
 \ENDIF
\ENDFOR
\ENSURE $\fat \eta^{k+1}$
\end{algorithmic}
\caption{The proposed algorithm for obtaining the $(k+1)$-th sample from $\pi ( \fat \eta | \fat y,\fat \nu, \fat \theta) $ in the data-rich block. By choosing $I=1$, the sampling scheme introduced in  Section \ref{sec:datarich} is obtained. For $I\geq2$ the modified sampling scheme, which is derived in Appendix \ref{CondInd},  is obtained for the partitions. }
\end{algorithm} 

\subsection{Sampler for the data-poor block}
\label{sec:datapoor}
The parameters $(\fat \nu, \fat \theta )$ in the data-poor block are, by construction, conditionally independent of $\fat y$ conditioned on the vector $\fat \eta$ from the data-rich block, that is, $$\pi(\fat \nu, \fat \theta \mid \fat y, \fat \eta) = \pi(\fat \nu, \fat \theta \mid \fat \eta).$$ The conditional posterior density  function of the data-poor block is therefore invariant of the choice of likelihood function and proportional to
\begin{equation}
\label{PoorPost}
	\pi(\fat \nu, \fat \theta \mid \fat y, \fat \eta) \propto \	\pi(\fat \nu \mid \fat \eta, \fat \theta  ) \pi(\fat \theta)
\end{equation}
where the conditional density function $\pi(\fat \nu \mid \fat \eta, \fat \theta  )$ is a Gaussian density of the form given in equation (\ref{condilater}). Moreover, if the Gaussian density functions in the prior assumptions in (\ref{priors}) are GMRFs with sparse precision structures then the Gaussian density function in (\ref{condilater}) retains the sparse GMRF structure induced by the prior assumption, by known results about conditioning of GRMFs \citep{rue2005gaussian}. Fast sampling algorithms for GMRFs can thus be implemented to obtain samples from the Gaussian density function in (\ref{condilater}), as discussed in \cite{rue2001fast}.

The relation in (\ref{PoorPost}) and the Gaussianity of $\pi(\fat \nu \mid \fat \eta, \fat \theta  )$ in (\ref{condilater}) motivate the following Metropolis--Hastings based sampling algorithm, which is a modified version of the one block sampler of \cite{knorr2002block}. For some proposal density $q(\fat \theta^* \mid \fat \theta^k)$ for the hyperparameters $\fat \theta$, a new proposed value $(\fat \nu^*, \fat \theta^*)$ is generated jointly as follows: 
\begin{align}
\label{sampla1}
\fat \theta^* &\sim q(\fat \theta^* \mid \fat \theta^k) \\ \nonumber
\fat \nu ^* &\sim \pi (\fat \nu^* \mid \fat \eta^{k+1},\fat \theta^*).
\end{align}
Denote the proposal density implied by (\ref{sampla1}) with $q(\fat \nu^*, \fat \theta^*\mid\fat \nu^k, \fat \theta^k)$. The proposed value $(\fat \nu^*, \fat \theta^*)$ is then accepted jointly with acceptance probability 
\begin{align}\label{allratio}
  \alpha = \min\left\{1,\frac{\pi( \fat \nu^*, \fat \theta^*\mid \fat y, \fat \eta^{k+1}) }{\pi( \fat \nu^k, \fat \theta^k\mid \fat y, \fat \eta^{k+1})} \frac{q(\fat \nu^k, \fat \theta^k\mid\fat \nu^*, \fat \theta^*)}{q(\fat \nu^*, \fat \theta^*\mid\fat \nu^k, \fat \theta^k)}\right\}.
\end{align}
In Lemma \ref{maindatapoor} we show how the acceptance ratio in (\ref{allratio}) can be simplified, which is modified version of the results shown in \cite{knorr2002block}.  

\begin{lemma}
\label{maindatapoor}
Assume the proposal density implied by (\ref{sampla1}) for the data-poor block, and denote the proposal density with $q(\fat \nu^*, \fat \theta^*\mid\fat \nu^k, \fat \theta^k)$. The corresponding acceptance ratio in (\ref{allratio}), can be simplified to
\begin{align}
\label{testa}
&\frac{\pi( \fat \nu^*, \fat \theta^*\mid \fat y, \fat \eta^{k+1}) }{\pi( \fat \nu^k, \fat \theta^k\mid \fat y, \fat \eta^{k+1})}
\frac{q(\fat \nu^k, \fat \theta^k\mid\fat \nu^*, \fat \theta^*)}{q(\fat \nu^*, \fat \theta^*\mid\fat \nu^k, \fat \theta^k)} =\frac{\pi(\fat \theta^* \mid\fat\eta^{k+1}) }{\pi(\fat \theta^k \mid\fat\eta^{k+1}) }
   \frac{q(\fat \theta^k \mid \fat \theta^*)}{q(\fat \theta^* \mid \fat \theta^k)}
   \end{align}  
and is therefore independent of the value of $\fat \nu$.    
\end{lemma}

In other words, the acceptance ratio in (\ref{testa})  is only dependent on the acceptance ratio for $\fat \theta$. Further, since the conditional posterior $\pi(\fat \nu \mid \fat \eta, \fat \theta)$ is a known Gaussian the proposed sampling strategy scales well in terms of computational efficiency as the dimensions of the data-poor component of the latent field $\fat \nu$ increases. 



When the Gaussian models in the prior assumptions (\ref{priors}) are GMRF density functions with a sparse precision structure, the ratio in (\ref{testa}) is computationally costly  to calculate directly, since $\pi(\fat \theta \mid\fat\eta ) \propto \pi(\fat \theta) \pi(\fat \eta \mid \fat \theta)$ and $\pi(\fat \eta \mid \fat \theta)$ does not necessarily preserve the sparse GMRF structure. However, as  the ratio in (\ref{testa})  is only dependent on the acceptance ratio for $\fat \theta$ it can be shown that the ratio in (\ref{testa}) can be rewritten in order to preserve the sparse GMRF precision structure, as stated in the Theorem \ref{maindatapoor2}.

\begin{theorem}
\label{maindatapoor2}  The term ${\pi(\fat \theta^* \mid\fat\eta^{k+1}) }\big/{\pi(\fat \theta^k \mid\fat\eta^{k+1}) }$ in (\ref{testa}) can be rewritten as
\begin{align}\label{gmrfratio00}
&\frac{\pi(\fat \theta^* \mid\fat\eta^{k+1}) }{\pi(\fat \theta^k \mid\fat\eta^{k+1}) } =  \frac{\pi (\fat \theta ^*)}{\pi(\fat \theta ^k)}\cdot
  \frac{\pi( \fat \eta^{k+1} \mid \fat 0, \fat \theta^*) \pi( \fat 0 \mid \fat \theta^*) }{\pi( \fat 0 \mid\fat \eta^{k+1}, \fat \theta^*) }\cdot
  \frac{\pi( \fat 0 \mid\fat \eta^{k+1}, \fat \theta^k) }{\pi( \fat \eta^{k+1} \mid \fat 0, \fat \theta^k) \pi( \fat 0 \mid \fat \theta^k) }
\end{align} 
Additionally, the conditional density functions on the right hand side in (\ref{gmrfratio00}) on a logarithmic scale are
\begin{align}
\label{reiknitrikk}
	\log \pi(\fat \eta | \fat 0, \fat \theta) &= \frac 1 2 \log \det \fat Q_{\epsilon}  
	- \frac 1 2 \fat \eta\trp \fat Q_{\epsilon}  \fat \eta + \text{const} \nonumber \\ 
\log \pi(\fat 0 | \fat \theta) &= \frac 1 2 \log \det \fat Q_{\nu} + \text{const} \\
\log \pi(\fat 0 | \fat \eta, \fat \theta) &= \frac 1 2 \log \det\left( \fat Q_{\nu} + \fat Z\trp \fat Q_{\epsilon} \fat Z \right) + \nonumber \\
	&\left( \left( \fat Q_{\nu} + \fat Z\trp \fat Q_{\epsilon} \fat Z \right)\inv \fat Z\trp \fat Q_{\epsilon}\fat\eta\right)\fat Z\trp\fat Q_{\epsilon} \fat \eta  + \text{const}\nonumber
\end{align}
Moreover, if the Gaussian prior density functions in  (\ref{priors}) are GMRFs with sparse precision structures, then all of the conditional density functions on the right hand side of (\ref{gmrfratio00}) are GMRFs with sparse precision structures.

\end{theorem}
Theorem \ref{maindatapoor2} shows how the ratio in (\ref{allratio}) can be calculated with low computational cost by using the results in (\ref{testa}), (\ref{gmrfratio00}) and (\ref{reiknitrikk}) in case of GMRFs with sparse precision structures. This is a key result for the implementation of the proposed sampling scheme in the data-poor block for GMRFs with sparse precision structures. The algorithm for the sampling scheme in the data-poor block is is summarized in Algorithm 2.

\begin{algorithm}[htp]
\label{alg_2}
\caption{The proposed algorithm for obtaining the $(k+1)$-th sample from $\pi (\fat \nu, \fat \theta  | \fat y, \fat \eta ) $ in the data-poor block. }
\begin{algorithmic}[1]
\REQUIRE $(\fat \nu^k, \fat \theta^k, \fat \eta^{k+1})$
\STATE  Sample each element of $\fat \theta^*$ from a proposal density $q(\fat \theta^* \mid \fat \theta^k)$
\STATE Calculate 
\begin{align*}
	r = \frac{\pi (\fat \theta ^*)}{\pi(\fat \theta ^k)}&\cdot\frac{\pi( \fat \eta^{k+1} | \fat 0, \fat \theta^*) \pi( \fat 0 | \fat \theta^*) }{\pi( \fat 0 |\fat \eta^{k+1}, \fat \theta^*) }  \frac{\pi( \fat 0 |\fat \eta^{k+1}, \fat \theta^k) }{\pi( \fat \eta^{k+1} | \fat 0, \fat \theta^k) \pi( \fat 0 | \fat \theta^k) }
	\cdot 
	 \frac{q(\fat \theta^k \mid \fat \theta^*)}{q(\fat \theta^* \mid \fat \theta^k)} 
\end{align*}
on a logarithmic scale, using the equations in (\ref{reiknitrikk}) for the conditional posterior densities functions
\STATE Calculate $\alpha  = \min\left\{ 1, r \right\}$

\STATE Sample $u \sim \mathcal{U}(0,1)$

\IF{$\alpha > u$}
\STATE Calcualte $\fat Q_{\nu|\eta} = \fat{Q}_\nu+\fat Z\trp\fat{Q}_\epsilon\fat Z$
\STATE Sample $\fat \nu ^*$ from
\begin{align*}
	\fat \nu^* | \fat \eta^{k+1},\fat \theta^{*} \sim  \mathcal N  \left( \fat\nu^* \Big | \fat Q_{\nu|\eta} ^{-1}(\fat Q_\nu \fat \mu_\nu + \fat Z\trp \fat Q_\epsilon \fat\eta^{k+1}) , \fat Q_{\nu|\eta} ^{-1}\right)
\end{align*}
    \STATE $(\fat \nu^{k+1}, \fat \theta^{k+1})= (\fat \nu^*, \fat \theta^*)$
  \ELSIF{$\alpha < u$}
  \STATE $(\fat \nu^{k+1}, \fat \theta^{k+1})  = (\fat \nu^k, \fat \theta^k)$ 
 \ENDIF
\ENSURE $(\fat \nu^{k+1}, \fat \theta^{k+1})$
\end{algorithmic}
\end{algorithm}

\section{Examples}
\label{sec:examples}
Two examples are presented in this section where the MCMC split sampler is applied to obtain posterior samples from the proposed models.
In the former example, a data set on annual mean precipitation in Iceland is modeled with a LGM that has a spatial model structure at the latent level. The latter example is on extreme flood events. 

We will emphasize that the aim of this section is to present some of the possibilities offered by the MCMC split sampler rather than to claim which model is the best for each data set. The main purpose of the first example is to demonstrate that the MCMC split sampler is well suited to infer LGMs with a spatial models on both location and scale parameters of the data density function, and that the computational efficiency of the sampler scales well as the number of unobserved spatial grid points increases. The main goal of the latter example is show that the MCMC split sampler is designed to infer LGMs with a non-Gaussian three parameter data density function, where all the three parameters are modeled with latent Gaussian models.

\subsection{Annual mean precipitation in Iceland}
\label{sec:rain}
\begin{figure} [htb]
	\centering
     	\includegraphics[width=1\linewidth]{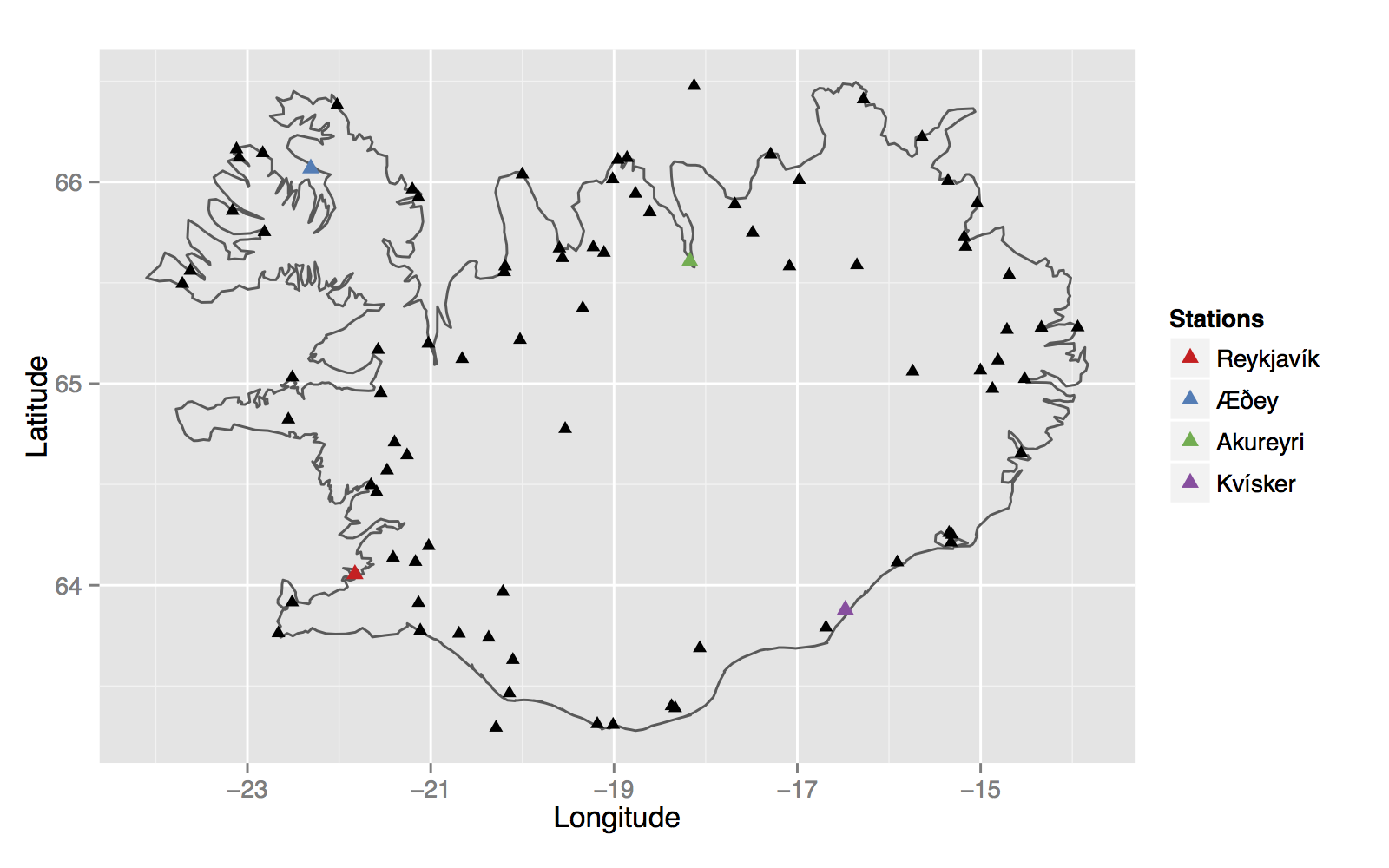}
	\caption{The $I=86$ observational sites in Iceland. Reykjavík is marked with red, Æðey is marked with blue, Akureyri is marked with green and Kvísker is marked with purple. }
	\label{obssites}
\end{figure}

\begin{figure} [htb]
	\centering
     	\includegraphics[width=1\linewidth]{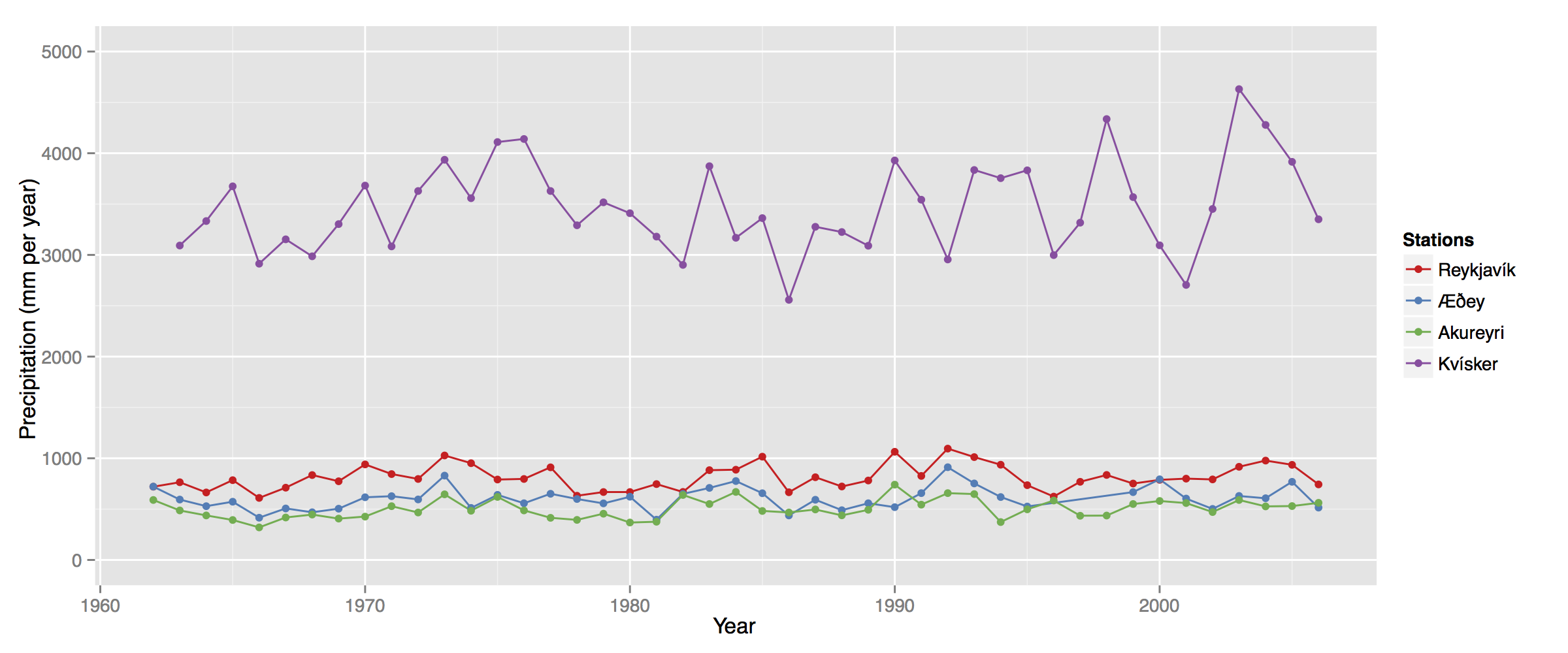}
	\caption{Times series over the years 1962 to 2006 on annual precipitation. The time series are based on observations from Reykjavík (red curve), Æðey (blue curve), Akueyri (green curve) and Kvísker (purple curve).} 
	\label{time1}
\end{figure}

The data set analyzed in this section is on observations on annual precipitation from 86 observational sites across Iceland, see Figure \ref{obssites}, over the years 1962 to 2006. Times series on annual  precipitation from the observational sites Reykjavík, Æðey, Akureyri and Kvísker are shown in Figure \ref{time1}. The data was provided by the Icelandic Meteorological Office (IMO).

 A LGM with a SPDE spatial model structure \citep{lindgren2011explicit} at the latent level is presented to obtain the spatially varying distributional properties of annual precipitation over the domain. We will demonstrate that the computational efficiency of the MCMC split sampler scales well as the number of grid points in the mesh in the SPDE approach increase.

\subsubsection*{Model setup}
\label{Model1}

\paragraph{The data level:} The data were modeled with a LGM assuming the Gaussian distribution for the observations and conditional independence over the observational sites. That is, let $y_{it}$ denote the annual precipitation at observational site $i$ at year $t$ then the data density function becomes
\begin{align*}
	&\pi(y_{it} | \mu_i, \tau_i)  = \\
	&\mathcal N \left(y_{it} \mid \mu_{i}, \exp(\tau_i) \right), \quad i=1,\ldots,I, \quad t=1,\ldots, T
\end{align*}
where $I$ is the number of sites, $T$ is the number of years; $\mu_{i}$ and $\tau_i$  are mean and log-variance parameters, respectively, which are both allowed to vary spatially.

\paragraph{The latent level:} The following model structure was implemented for the mean parameter  $\fat \mu = (\mu_1,\ldots, \mu_I)$ at the latent level of the model,
\[
	\fat \mu = \fat X_\mu \fat \beta_\mu + \fat A_\mu \fat u_\mu +  \fat{\epsilon}_\mu,
\]
where $\fat X_\mu$ is a design matrix consisting of a vector of ones and  covariates that are based on the meteorological model of \cite{crochet2007estimating}, see \cite{geirsson2015computationally} for details; $\fat{\beta}_\mu$ are the corresponding weights; $\fat u_\mu$ denotes a Mat\'ern type spatial field constructed with the SPDE approach \citep{lindgren2011explicit} on a triangulated mesh over the spatial domain with a smoothness parameters chosen as one, which corresponds to an almost once differentiable Matérn field and $\alpha=2$ in the SPDE method; $\fat A_{\mathcal{S}}$ is a known projection matrix; the matrix product $\fat A_{\mathcal{S}}\fat u_\mu $ then denotes the spatial effect at the observational sites, which captures the spatial variation in the data that is unexplained by the covariate and $ \fat{\epsilon}_\mu$ is an unstructured random effect. Analogous model structure is also implemented for the log-variance parameter, that is
\[
	\fat \tau = \fat X_\tau \fat \beta_\tau + \fat A_{\mathcal{S}} \fat u_\tau + \fat \epsilon_\tau.
\]
where $\fat X_{\tau}$ is a design matrix consisting of a vector of ones and the aforementioned meteorological covariate on a logarithmic scale. 

Working within the LGM setup, the following prior density functions were assigned to parameters at the latent level of the model.
\begin{align*}
\label{prior}
	\pi(\fat \beta_\mu) &= \mathcal{N}(\fat \beta_\mu \mid \fat 0, \kappa_{\beta \mu}^{-1} \fat I),  &  \pi(\fat \beta_\tau) &= \mathcal{N}(\fat \beta_\tau \mid \fat 0, \kappa_{\beta \tau}^{-1} \fat I),  \nonumber \\
	\pi(\fat u_\mu) &= \mathcal{N}(\fat u_\mu \mid \fat 0 , \fat Q^{-1}_{ u \mu}), & \pi(\fat u_\tau) &= \mathcal{N}(\fat u_\tau \mid \fat 0 , \fat Q^{-1}_{u \tau}),\\
	\pi(\fat \epsilon_\mu) &= \mathcal{N}(\fat \epsilon_\mu \mid \fat 0, \sigma_{\epsilon \mu}^{2} \fat {I}), & \pi(\fat \epsilon_\tau) &= \mathcal{N}(\fat \epsilon_\tau \mid \fat 0, \sigma_{\epsilon \tau }^{2} \fat {I}). \nonumber
\end{align*}
The parameter values $\kappa_{\beta \mu} = 0.0025$ and $\kappa_{\beta \tau}= 0.25$ were fixed in the prior distributions for $\fat \beta_\mu$ and $\fat \beta_\tau$. The precision matrices $ \fat Q_{ u \mu}$ and  $\fat Q_{ u \mu}$ are constructed with SPDE approach, and have sparse GMRF precision structures. Further, the precision matrix $ \fat Q_{ u \mu}$  has two parameters, $\sigma_{u \mu}$ and  $\kappa_{u \mu}$, which serve as hyperparameters of the spatial model for $\mu$. The hyperparameters $\sigma_{u \mu}$ and  $\kappa_{u \mu}$ are related to the marginal variance and range of the spatial field, respectively. Analogous structure holds for $ \fat Q_{ u \tau}$. The parameters $\sigma_{\epsilon \mu}^{2}$ and $\sigma_{\epsilon \tau }^{2}$ are unknown variance parameters for the unstructured random effects.

\paragraph{The hyper level:} Let $\fat \theta$ denote all the hyper parameters of the model that are not fixed, that is
\[
	\fat \theta = (\sigma_{u \mu}, \kappa_{u \mu},  \sigma_{\epsilon \mu}, \sigma_{u \tau}, \kappa_{u \tau},  \sigma_{\epsilon \tau}).
\]
Lognormal prior distributions with fixed parameters were assigned to the hyperparmeters in $\fat \theta$.

\subsubsection*{Posterior inference}

In order to apply the MCMC split sampler to the aforementioned model, the model parameters are assigned to the data-rich block which includes $\fat \eta=(\fat\mu, \fat\tau)$ and the data-poor block which consists of $\fat \nu=(\fat\beta_\mu, \fat u_\mu,\fat \beta_\tau , \fat u_\tau)$ and the hyperparameters $\fat \theta$.

The aforementioned model setup and prior assumptions are equivalent to the setup implied in equations (\ref{mform}) and  (\ref{priors}) with 
\begin{align}
\label{Z1}
	&\fat Z = \begin{pmatrix} 
		\fat X_\mu & \fat A_\mu & \cdot & \cdot  \\
		\cdot & \cdot & \fat X_\tau & \fat A_\tau  \\
	\end{pmatrix}
	\text{, }
	\quad
	\fat Q_\epsilon = \begin{pmatrix} 
		\sigma^{-2}_{\epsilon \mu} \fat I  & \cdot  \\
		\cdot & \sigma^{-2}_{\epsilon \tau} \fat I  \\
	\end{pmatrix} \nonumber\\
	&\fat Q_\nu = \begin{pmatrix} 
		\kappa_{\beta \mu}  \fat I  & \cdot & \cdot & \cdot \\
		\cdot  &  \fat Q_{ u \mu} & \cdot & \cdot \\
		\cdot  & \cdot & \kappa_{\beta \tau}  \fat I   & \cdot \\
		\cdot  & \cdot & \cdot  &  \fat Q_{ u \tau}\\
	\end{pmatrix}.
\end{align}

\paragraph{Data-rich block: } The conditional posterior $\pi(\fat \eta \mid \fat y, \fat \nu, \fat \theta)$ in the data-rich block is intractable. However, the logarithm of the conditional posterior of the data-rich block is of the same form as in equation  (\ref{logdens1}), with $\fat Z$ and $\fat Q_\epsilon$ defined in equation (\ref{Z1}) and 
\begin{align}
\label{ff1}
	f (\fat \eta )  = \sum_{i=1}^I f_i(\fat \eta_{i}) = \sum_{i=1}^I \sum _{t \in \mathcal{A}_i} \log \mathcal N (y_{it} | \mu_i, \exp{\tau_i}),
\end{align}
where the set $\mathcal{A}_i$ contains the indices of the years $t$ observed at site $i$. By model assumptions, the vectors $\fat \eta_{i}=(\mu_i, \tau_i)$
become conditionally independent  in the conditional posterior \mbox{$\pi ( \fat \eta \mid \fat y, \fat \nu, \fat \theta) $} over observational sites $i$. 
This demonstrates that the modification of the sampling scheme in Section \ref{sec:datarich}, outlined in Appendix \ref{CondInd}, is applicable. Therefore Algorithm 1 was used to obtain MCMC samples from the conditional posterior $\pi(\fat \eta \mid \fat y, \fat \nu, \fat \theta)$ with $I=86$,  $\fat Q_\epsilon$  as in (\ref{Z1}) and $f(\fat \eta)$ as in (\ref{ff1}).\\

\paragraph{Data-poor block:} In order to implement the sampling strategy outlined in Section \ref{sec:datapoor} and to obtain MCMC samples from the conditional posterior $\pi(\fat \nu, \fat \theta \mid \fat y, \fat \eta)$, a proposal density $q$ for the hyperparameters $\fat \theta$ must be chosen. In this example, the proposal strategy suggested in \citep{knorr2002block} is used for each element of $\fat \theta$. That is, let $\theta_i^* = f\theta_i^k$ where the scaling factor $f$ has the density
\begin{equation}
\label{HavardProp}
	\pi(f) \propto 1 + 1/f  \quad \text{for $f \in [1/F, F]$}
\end{equation}
where $F>1$ is a tuning parameter. \cite{knorr2002block} show that this is a symmetric proposal density in the sense that $q(\theta_i^*|\theta_i^k) = q(\theta_i^k|\theta^*_i)$. Therefore, by using this proposal density, the acceptance probability in equation (\ref{allratio}) simplifies to
\begin{align*}
  \alpha = \min\left\{1,\frac{\pi(\fat \theta^* \mid\fat\eta^{k+1}) }{\pi(\fat \theta^k \mid\fat\eta^{k+1}) }\right\}
\end{align*}
Moreover, the ratio in (\ref{gmrfratio00}) in Theorem \ref{maindatapoor2} was used to calculate the acceptance probability, which preserves the sparse GMRF precision structure induced by the SPDE approach.  Thus, Algorithm 2 is was implemented to obtain MCMC samples from the conditional posterior \mbox{$\pi ( \fat \eta \mid \fat y, \fat \nu, \fat \theta) $}, with $\fat Z$, $\fat Q_\epsilon$ and $\fat Q_\nu$ defined in (\ref{Z1}) and the proposal density in (\ref{HavardProp}).

\subsubsection*{Convergence diagnostics}
\begin{figure} [h!]
	\centering
     	\includegraphics[width=0.90 \linewidth]{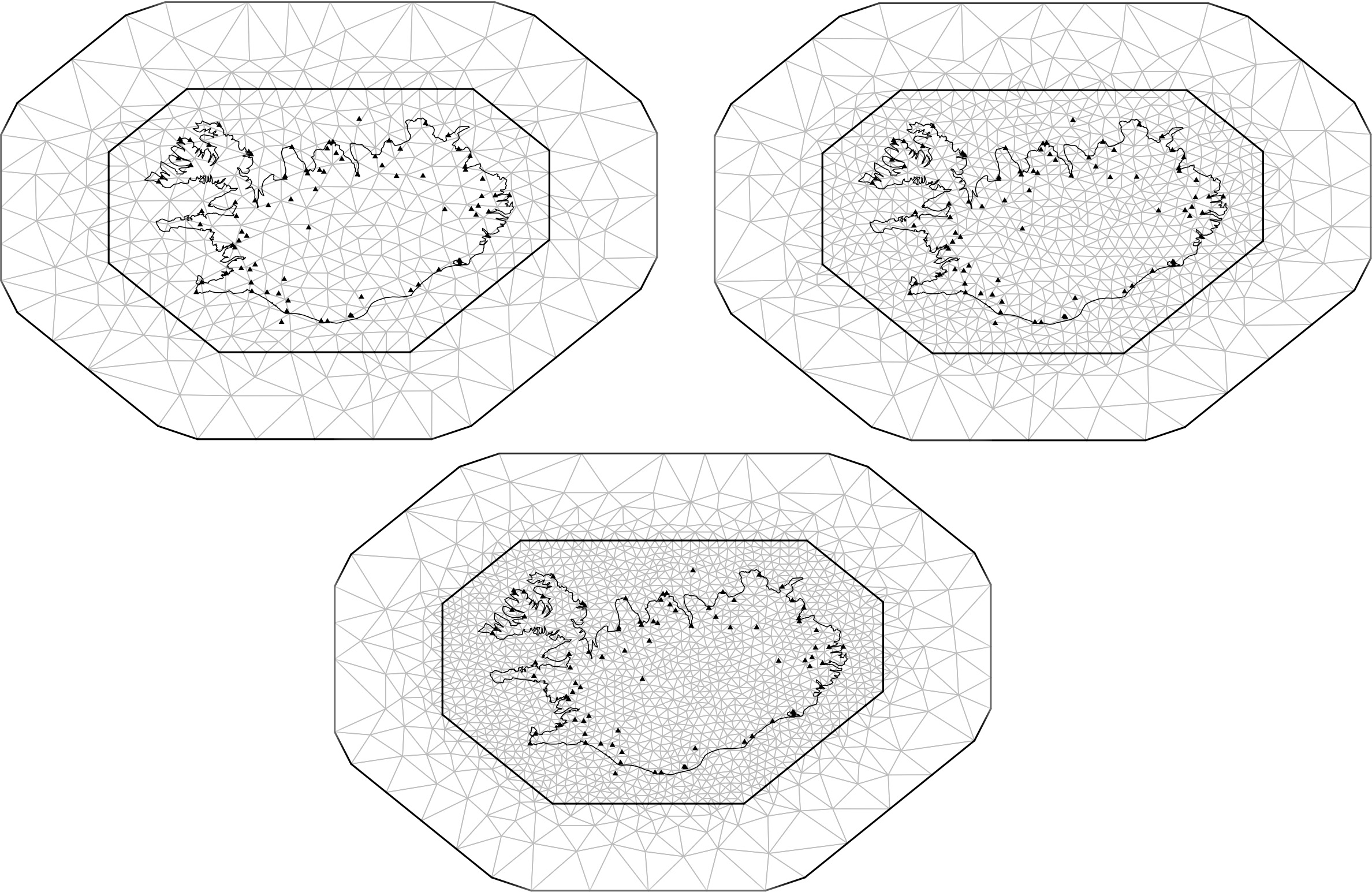}
	\caption{The triangulated meshes over the spatial domain, based on the coarse mesh (top left), medium mesh (top right) and dense mesh (bottom). }
	\label{MeshAll}
\end{figure}

The following convergence diagnostics are based on four MCMC chains sampled in parallel with the MCMC split sampler from the proposed model. Each chain was calculated with 50000 iterations where 10000 iterations were burned in. The posterior inference was carried out separately for three different mesh resolutions. That is, a coarse resolution based on 411 mesh points; a medium resolution based on 858 mesh points; and a dense resolution based 1752 mesh points. In Figure \ref{MeshAll} the three different meshes are presented on a same scale. The top, middle and bottom panels in Figure \ref{MeshAll} show the coarse resolution, medium resolution and dense resolution meshes, respectively. Runtime, on a modern desktop (Ivy Bridge Intel Core i7-3770K, 16GB RAM and a solid state hard drive), was approximately 6, 6.5 and 7 hours for the coarse, medium and dense mesh resolution, respectively.  All calculations were carried out using $\mathtt R$.

\begin{figure} [htb]
	\centering
     	\includegraphics[width=0.85\linewidth]{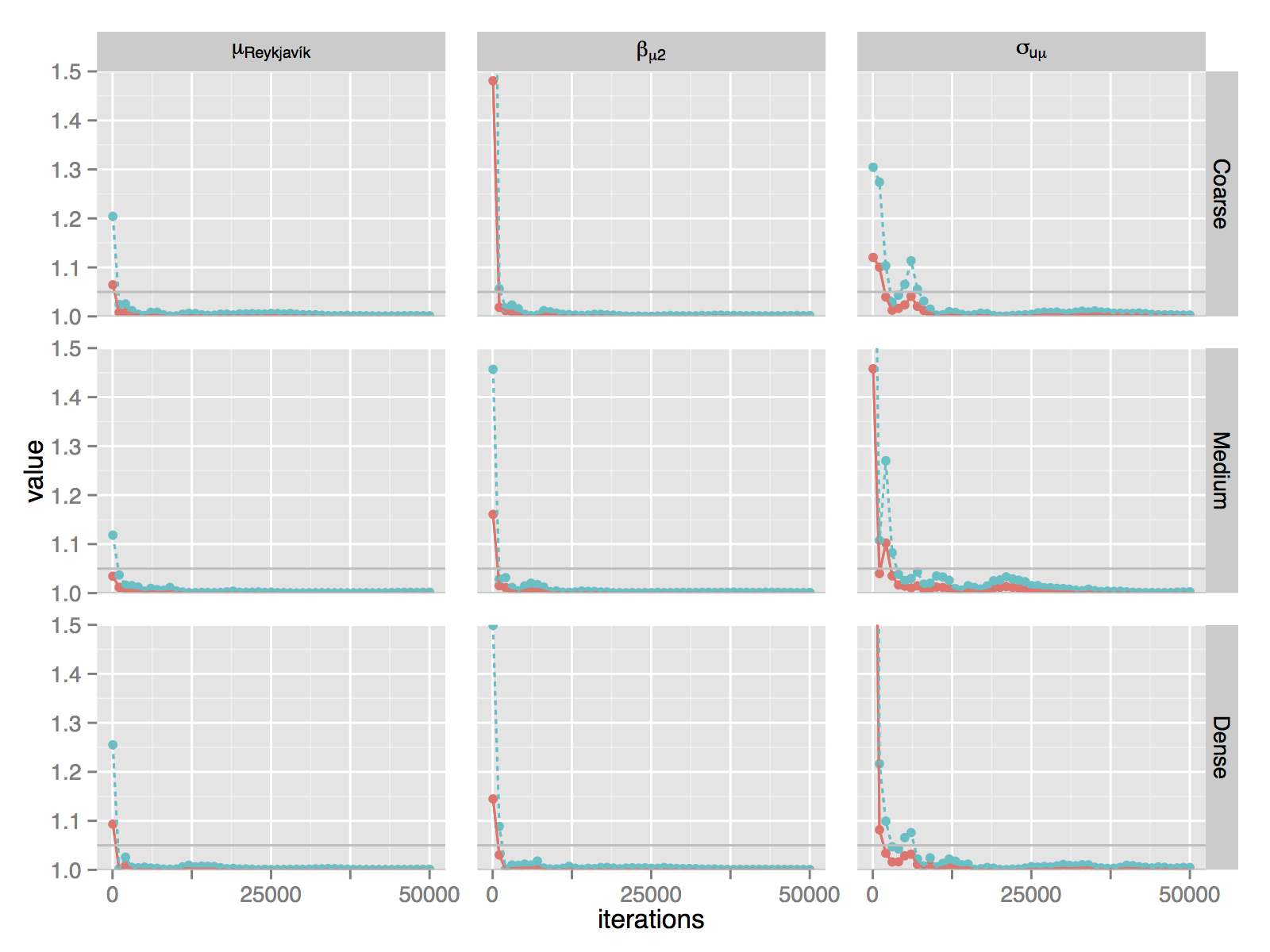}
	\caption{Gelman--Rubin plots for $\mu$, $\beta_{\mu 2}$, and $\sigma_{u \mu}$ for three different mesh resolutions. The red solid line denotes the median of the Gelman--Rubin statistics, and the green dashed line denoted the  upper limit of the 95\% confidence interval for the Gelman--Rubin statistics. The first row is based on a coarse resolution (411 mesh points). The second row is based on a medium resolution (858 mesh points). The third row is based on a dense resolution (1752 mesh points). The results demonstrate the MCMC split sampler has converged in the mean after roughly 7500 iterations.}
	\label{GM_1}
\end{figure}

Gelman--Rubin plots, based on the MCMC runs, of the the mean parameter $\mu$ in Reykjavík; the covariate coefficient  $\beta_{\mu 2}$; and the marginal standard deviation for the spatial field $\sigma_{u \mu}$ are shown in the first, second  and third column, respectively, in Figure \ref{GM_1}. The results based on the coarse resolution, medium resolution and dense resolution meshes for the aforementioned parameters are shown in the first, second and third row, respectively, in Figure \ref{GM_1}. A comparison of the results in Figure \ref{GM_1} between the different mesh resolutions reveals that the convergence in the mean is achieved for the the three parameters at a similar rate. Furthermore, the Gelman--Rubin plots in Figure \ref{GM_1} show that the sampler has converged in the mean after roughly 7500 iterations for all mesh resolutions. Similar results hold for all of the other model parameters (results not shown).

\begin{figure} [htb]
	\centering
     	\includegraphics[width=0.85\linewidth]{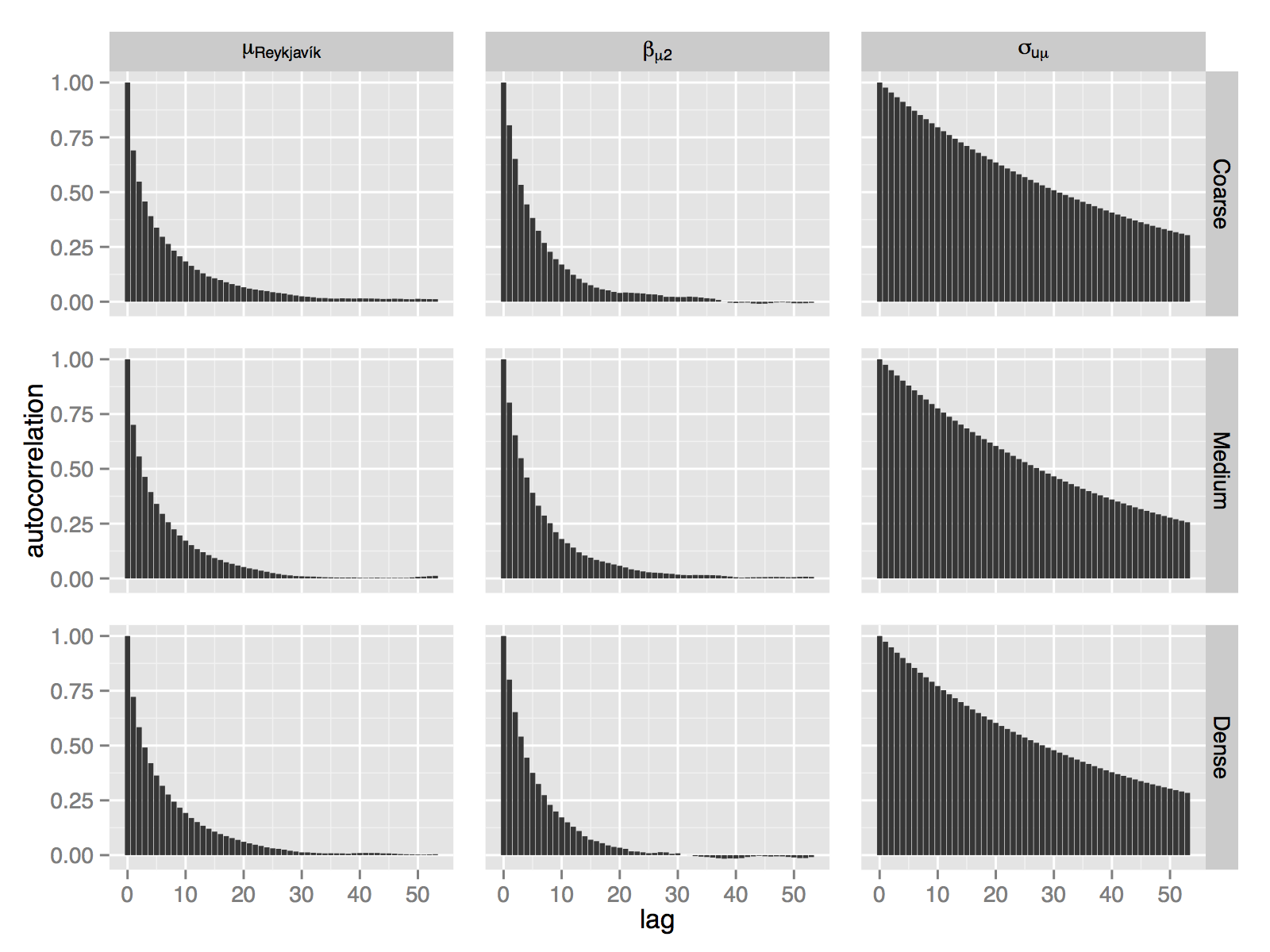}
	\caption{Autocorrelation plots for $\mu$, $\beta_{\mu 2}$, and $\sigma_{u \mu}$ for three different mesh sizes. The first row is based on a coarse resolution (411 mesh points). The second row is based on a medium resolution (858 mesh points). The third row is based on a dense resolution (1752 mesh points). The results demonstrate that the autocorrelation in the MCMC chains decays rapidly and in invariant of the number of points in the mesh.}
	\label{Auto_1}
\end{figure}

Autocorrelation plots for the same set of parameters and arranged identically as in Figure \ref{GM_1} are shown in Figure \ref{Auto_1}. The results demonstrate that the MCMC chains for the mean parameter $\mu$ in Reykjavík and the covariate coefficient  $\beta_{\mu 2}$ exhibit a negligible autocorrelation after lag 10. The MCMC samples of the hyperparameter  $\sigma_{u \mu}$ show autocorrelation around 0.3 at lag 50. Similar results hold for all the other model parameters (results not shown). 

Relying on the Gelman-Rubin statistics and the autocorrelation plots, the MCMC chains exhibit all signs of having converged. Moreover, the autocorrelation plots in Figure \ref{Auto_1} reveal that the autocorrelation in the MCMC chains does not increase with number of mesh points, which in turn indicates that the autocorrelation in the MCMC chains is invariant of the dimensions of the data-poor part of the latent field $\fat \nu$. These results demonstrate that the MCMC split sampler retains its computational efficiency when the number of mesh points increases, which is to be expected as the acceptance probability in (\ref{testa}) in the data-poor block in independent of the  $\fat \nu$.

Furthermore, as the acceptance probability in (\ref{testa}) within the data-poor block is only dependent on the hyperparameters, the autocorrelation seen in MCMC chains for the hyperparameter $\sigma_{u \mu}$ in Figure \ref{Auto_1} is mainly affected by the choice of proposal density for $\fat \theta$, which is in this example the sampler in (\ref{HavardProp}). In Section \ref{Flood} we will demonstrate the modularity of the MCMC split sampler, by choosing another proposal density for the hyperparmeters  $\fat \theta$ in Algorithm 2 which significantly reduces the autocorrelation in MCMC chains for $\fat \theta$. 

\subsection{Flood analyzis}
\label{Flood}
In this section, we present a simulation study on extreme events. The data set consists of simulations of monthly maximum instantaneous flow  based on characteristics of ten river catchments around Iceland. The characterizing features that were used to simulate the data for each river were chosen as river catchment area and maximum daily precipitation, as both river catchment area and maximum daily precipitation are known to be positively correlated with maximum instantaneous flow, see \cite{davidhsson2015bayesian} and \cite{crochet2012estimating}. The simulated time series were chosen to be 150 years.

\subsubsection*{Model setup}

\paragraph{The data level:} The data were modeled with a LGM assuming the generalized extreme value distribution (g.e.v.) for the observations. To that extend, let $y_{mj, t}$ denote the value from river $j$ at month $m$ and year $t$, with a cumulative density function of the form
\begin{equation*}
	F(y_{mj, t}) =  \exp\left\{ - \left(1+ \xi_{mj}\left(\frac{y_{mj, t}-\mu_{mj}}{\sigma_{mj}}\right)\right)^{-1/\xi_{mj}} \right\}
\end{equation*}
if  $1+\xi_{mj}(x-\mu_{mj)}/\sigma_i>0$, $F(y_{it}) = 0$ otherwise. The parameters $\mu_{mj}$, $\sigma_{mj} $ and $\xi_{mj}$ are the location, scale and shape parameters of the g.e.v. distribution for river $j$ in month $m$. Additionally,  $J$ is the number of rivers and $T$ is the number of years.  Further, the data is assumed independent between both rivers and between months. These assumptions were made for demonstrative purposes. 

\paragraph{The latent level:} The location and scale parameters are modeled on a logarithmic scale at the latent level, which is modelling setup along the lines presented in \cite{cuanne1971} and \cite{grehy1996presentation}. Thus, define $\lambda_{mj} = \log \mu_{mj}$ and $\tau_{mj} = \log \sigma_{mj}$. The shape parameter is modeled on its native scale. 

As discussed in \cite{davidhsson2015bayesian}, the underlying processes of monthly maximum instantaneous flow exhibit a seasonal behavior. Therefore, the following seasonal model is proposed for the location parameter on a logarithmic scale. That is, 
\begin{align}
\label{seso}
	\lambda_{mj} &= \beta_{0,\lambda}  + u_{0,m,\lambda} +x_{1,mj}(\beta_{1,\lambda} + u_{1,m,\lambda})  \nonumber \\
	&+\ldots  +  x_{p,m,j}(\beta_{p,\lambda} + u_{p,m,\lambda}) + \epsilon_{mj,\lambda}
\end{align}
where $\beta_{0,\lambda}$ denotes an overall intercept term; $x_{i,mj}$ denotes the $i$-th covariate in month $m$ at the $j$-th river; $\beta_{i,\lambda}$ denotes the weight of the $i$-th covariate for $i=1,\ldots,p$; $u_{0,m,\lambda}$ denotes the seasonal random effect of the $m$-th month; $u_{i,m,\lambda}$ denotes the seasonal additional weight of the $i$-th covariate within month $m$; and $\epsilon_{mj,\lambda}$ denotes an unstructured random effect.

In order to write the model in a matrix form for the implementation of the MCMC split sampler, combine the location parameters for river $j$ over months. That is,
\[
	\fat \lambda_{j} = (\lambda_{1j},\ldots,\lambda_{12j})\trp, \quad j=1,\ldots, J
\]
and define the following 
\[
	\fat u_{i,\lambda} = (u_{i,1,\lambda},\ldots,u_{i,12,\lambda})\trp,  \fat A_{i,j} = \text{diag}( x_{i,j1}, \ldots, x_{i,j12}), 
\]
where $i=0,\ldots,p$ and $x_{0,jm} = 1$ denotes the intercept term for river $j$ and month $m$. Additionally, define
\[
	\fat X_j = \begin{pmatrix} 
		1 & x_{1,j,1}& \ldots & x_{p,j,1} \\
		1 & x_{1,j,2}& \ldots & x_{p,1,2} \\
		\vdots  && \vdots \\
		1 & x_{1,j,12}& \ldots& x_{p,1,12} \\
	\end{pmatrix},  \quad
	\fat A_j = \begin{pmatrix} 
	\fat A_{0,j}, \ldots, \fat A_{p,j}
	\end{pmatrix}.
\]	
%
%
%
The seasonal model presented in (\ref{seso}) for the log-location parameter for river $j$ can be written in matrix form as
\begin{align*}
	\fat \lambda_j = \fat X_j \fat \beta_\lambda + \fat A_j \fat u_\lambda \fat + \fat \epsilon_{j,\lambda}
\end{align*}
where $\fat \beta_\lambda = (\beta_{0,\lambda}, \ldots, \beta_{p,\lambda})\trp$, $\fat u_\lambda = (\fat u_{0,\lambda}, \ldots, \fat u_{p,\lambda})\trp$, and $\fat\epsilon_{j,\lambda} = (\epsilon_{1j,\lambda},\ldots,\epsilon_{12j,\lambda})$. By combing the seasonal model over rivers, the following holds

\begin{align*}
	\fat \lambda = \fat X \fat \beta_\lambda + \fat A \fat u_\lambda \fat + \fat \epsilon_\lambda
\end{align*}
where
\[
	\fat \lambda =
	\begin{pmatrix} 
	\fat \lambda_1 \\
	\vdots \\
	\fat \lambda_J
	\end{pmatrix}, 
	\fat X =
	\begin{pmatrix} 
	\fat X_1 \\
	\vdots \\
	\fat X_J
	\end{pmatrix},
	\fat A =
	\begin{pmatrix} 
	\fat A_1 \\
	\vdots \\
	\fat A_J
	\end{pmatrix},
	\fat \epsilon_\lambda =
	\begin{pmatrix} 
	\fat \epsilon_{1,\lambda}\\
	\vdots \\
	\fat \epsilon_{J,\lambda}
	\end{pmatrix}.
\]
Analogous model structure was also implemented for the log-scale parameter. That is, 
\begin{align*}
	\fat \tau = \fat X \fat \beta_\tau + \fat A \fat u_\tau + \fat \epsilon_\tau.
\end{align*}
A reduced model with a similar structure was implemented for the shape parameter $\xi$. That is 
\begin{align}
\label{seso2}
	\xi_{mj}  = \beta_{0,\xi}  + u_{0,m,\xi} + \epsilon_{mj,\xi}
\end{align}
where $\beta_{0,\xi}$ denotes an overall intercept term;  $u_{0,m,\xi}$ denotes the seasonal random effect of the $m$-th month; and $\epsilon_{mj,\xi}$ denotes an unstructured random effect. The full matrix model for $\xi$ becomes 
\begin{align*}
	\fat \xi = \fat 1_{12 J} \beta_{0,\xi} + \left(\fat 1_J \otimes \fat I_{12}\right) \fat u_\xi + \fat \epsilon_\xi
\end{align*}
where $\fat 1_n$ denoted an $n$-dimensional vector of ones. 

Working within the LGM framework, the following prior density functions were assigned to the latent parameters. First assign,

\begin{align*}
\pi(\fat \beta_\lambda) &= \mathcal{N}(\fat \beta_\lambda \mid \fat 0, \sigma_{\beta \lambda}^{2} \fat I),  \quad  \pi(\fat \beta_\tau) = \mathcal{N}(\fat \beta_\tau \mid \fat 0, \sigma_{\beta \tau}^{2} \fat I) \quad \pi(\beta_\xi) &= \mathcal{N}( \beta_\xi \mid 0, \sigma_{\beta \xi}^{2}).
\end{align*}
The parameters  $\fat \beta_\lambda, \fat \beta_\tau$ and $\beta_\xi$ are assumed \emph{a priori} to have a low precision on their native scales in order to let the data play the dominate role in their inference. Thus, the parameter values $\sigma_{\beta \lambda} = 4$, $\sigma_{\beta \tau}= 4$ and $\sigma_{\beta \xi}= 2$ were chosen for the prior density functions. 

Secondly, the selection of prior density functions for the seasonal random effects needs to incorporate a correlation structure that induces a strong correlation between neighbouring months. This is achieved by assigning the following prior density functions
\begin{align*}
	\pi(\fat u_\lambda)  &= \mathcal{N}(\fat u_\lambda \mid \fat 0, \text{diag}(\fat\psi_\lambda) \otimes \fat Q^{-1}_{u}), \\
	\pi(\fat u_\tau)  &= \mathcal{N}(\fat u_\tau \mid \fat 0, \text{diag}(\fat\psi_\tau) \otimes \fat Q^{-1}_u), \\
	\pi(\fat u_\xi)  &= \mathcal{N}(\fat u_\xi \mid \fat 0, \psi_\xi \fat Q_u^{-1})
\end{align*}
where $\fat\psi_\lambda = (\psi_{0,\lambda},\ldots,\psi_{p,\lambda})\trp$, $\fat\psi_\tau = (\psi_{0,\tau},\ldots,\psi_{p,\tau})\trp$ and $\psi_\xi$ serve as scaling parameters for the monthly random effects corresponding to the three intercepts and the covariates; and the $\fat Q_{u}(\kappa)$ is a $12\times 12$ circular band precision matrix that has the vector
\begin{align*}
[1 \quad -2(\kappa^2 + 2) \quad &\kappa^4 + 4\kappa^2 + 6 \quad -2(\kappa^2 + 2) \quad 1 ]
\end{align*}
on the diagonal band, as discussed in \cite{lindgren2011explicit}, which capture the autocorrelation between months. In this example, the decay parameters was fixed to simplify the inference and set equal to $\kappa=1$. Further, this value of $\kappa$ induces an  autocorrelation \emph{a priori}  between consecutive months. Third, for the unstructured random effects, the following priors were chosen. 
\begin{align*}
	\pi(\fat \epsilon_\lambda) &= \mathcal{N}(\fat \epsilon_\lambda \mid \fat 0, \sigma_{\epsilon \lambda}^{2} \fat {I}), \quad 
	\pi(\fat \epsilon_\tau) =\mathcal{N}(\fat \epsilon_\tau \mid \fat 0, \sigma_{\epsilon \tau }^{2} \fat {I}), \quad
	 \pi(\fat \epsilon_\xi) &= \mathcal{N}(\fat \epsilon_\xi \mid \fat 0, \sigma_{\epsilon \xi }^{2} \fat {I}). \nonumber
\end{align*}

\paragraph{The hyper level:} Let $\fat \theta$ denote all the hyperparameters of the model that are not fixed on a logarithmic scale for computational purposes. That is,
\begin{align*}
	\fat \theta = (&\log \psi_{0,\lambda}, \ldots, \log \psi_{p,\lambda}, \log \psi_{0,\tau}, \ldots, \log \psi_{p,\tau},\log \psi_\xi, \log \sigma_{\epsilon \lambda}^{2}, \log  \sigma_{\epsilon \tau}^{2}, \log  \sigma_{\epsilon \xi}^{2})
\end{align*}
Gaussian prior distributions with fixed parameters were assigned to the hyperparmeters in $\fat \theta$. 

\subsubsection*{Posterior inference}
The data-rich block includes $\fat \eta=(\fat\mu, \fat\tau, \fat \xi)$ and the data-poor block consists of $\fat \nu=(\fat\beta_\mu, \fat u_\mu,\fat \beta_\tau , \fat u_\tau, \beta_\xi, \fat u_\xi)$ and the hyperparameters $\fat \theta$.  For the implementations of the MCMC split sampler, define the following sparse matrices
\begin{align}
\label{ZZ2}
	\fat Z = \begin{pmatrix} 
		\fat X & \fat A & \cdot & \cdot  & \cdot & \cdot   \\
		\cdot & \cdot & \fat X & \fat A & \cdot & \cdot   \\
		\cdot & \cdot & \cdot & \cdot & \fat 1_{12 J} & \fat I_{12 J}  \\
	\end{pmatrix},
	\fat Q_\epsilon = \begin{pmatrix} 
			\sigma^{-2}_{\epsilon \lambda} \fat I & \cdot & \cdot \\
		\cdot & 	\sigma^{-2}_{\epsilon \tau } \fat I & \cdot \\
		 \cdot & \cdot &  \sigma^{-2}_{\epsilon \xi } \fat I
	\end{pmatrix}
\end{align}
and

\begin{align}
\label{QQ2}
\fat Q_\nu = \text{bdiag}\big(&\sigma_{\beta \lambda}^{-2} \fat I, \text{diag}(\fat\psi_\lambda) \otimes \fat Q^{-1}_{u},\sigma_{\beta \tau}^{-2} \fat I, \text{diag}(\fat\psi_\tau) \otimes \fat Q^{-1}_{u},  
\sigma_{\beta \xi}^{-2},\psi_\xi \fat Q^{-1}_u \big)
\end{align}
where {bdiag} denotes a block diagonal matrix. \\



\paragraph{Data-rich block:} The modified version of the sampling scheme in Section \ref{sec:datarich}, outlined in Appendix \ref{CondInd}, was used to obtain MCMC samples from the conditional posterior $\pi(\fat \eta \mid \fat y, \fat \nu, \fat \theta)$. The logarithm of the conditional posterior is of the same form as in equation (\ref{logdens1}), with $\fat Z$ and $\fat Q_\epsilon$ defined in equation (\ref{ZZ2}) and 
\begin{align}
\label{ff2}
	f(\fat \eta)  &= \sum_{m=1}^{12} \sum_{j=1}^{J} f_i(\fat \eta_{mj}) =  \sum_{m=1}^{12} \sum_{j=1}^{J} \sum _{t =1}^T \log \pi_{\text{gev}}\left(y_{mj, t} | \exp{\lambda_{mj}}, \exp{\tau_{mj}}, \xi_{mj}\right) 
\end{align}
where $\pi_{\text{gev}}$ denotes the density function of the generalized extreme value distribution. Therefore, Algorithm 1 was used to obtain MCMC samples from the conditional posterior from the data-rich block, with $I = J\cdot 12 = 120$, $\fat Q_\epsilon$ as in (\ref{ZZ2}) and $f(\fat \eta)$ as in (\ref{ff2}).\\

\paragraph{Data-poor block:} The sampling scheme outlined in Section \ref{sec:datapoor} was used to obtain MCMC samples from the conditional posterior $\pi(\fat \nu, \fat \theta \mid \fat y, \fat \eta)$ in the data-poor block. A proposal density based on the normal distribution centered on the last draw of $\fat \theta$, as discussed in \cite{roberts1997weak}, was selected for Algorithm 2, with a precision matrix $-c\fat H$ where $\fat H$ is a finite difference estimate of the Hessian matrix of $\log\pi(\fat \theta |\hat{\fat \eta})$ evaluated at the mode. That is, 
\begin{equation}
\label{HHH}
	\fat H \approx \nabla^2 \log \pi (\fat \theta| \hat{\fat \eta})  \big |_{\fat \theta = \fat \theta_0}
\end{equation}
where $\hat{\fat \eta}$ is the maximum likelihood estimate of $\fat \eta$ for each river and month; $\fat \theta_0$ is the mode of $\log\pi(\log \fat \theta |\hat{\fat \eta})$; and $c$ is a scaling constant. Conditioning on $\hat{\fat \eta}$, as opposed of $\fat \eta^{k+1}$ for example, removes the necessity to estimate $\fat H$ in every iteration. Moreover, setting a specific scaling $u$ removes the need for tuning. The scaling  $c = 2.382/\text{dim}(\fat \theta)$ was implemented, as it is optimal in a particular large dimension scenario, see \cite{roberts1997weak}. The resulting proposal density therefore becomes
\begin{equation}
\label{prop2}
	q(\fat \theta^* | \fat \theta^k) = \mathcal{N} \left( \fat \theta^* \mid \fat \theta^k, \left( -c\fat H \right)\inv \right).
\end{equation}
Algorithm 2 was thus implemented to obtain MCMC sampled from the conditional posterior within the data-poor block, with $\fat Z$, $\fat Q_\epsilon$ as in equation (\ref{ZZ2}); $\fat Q_\nu$ as in equation (\ref{QQ2}); and the proposal density in (\ref{prop2}).

\subsubsection*{Convergence diagnostics}
As in Section \ref{sec:rain}, the following convergence diagnostics are based on four MCMC chains sampled in parallel with the MCMC split sampler. Each chain was calculated with 50000 iterations where 10000 iterations were burned in. Runtime on the same desktop as in Section \ref{sec:rain} was approximately 7 hours.

\begin{figure} [hbt]
	\centering
     	\includegraphics[width=1\linewidth]{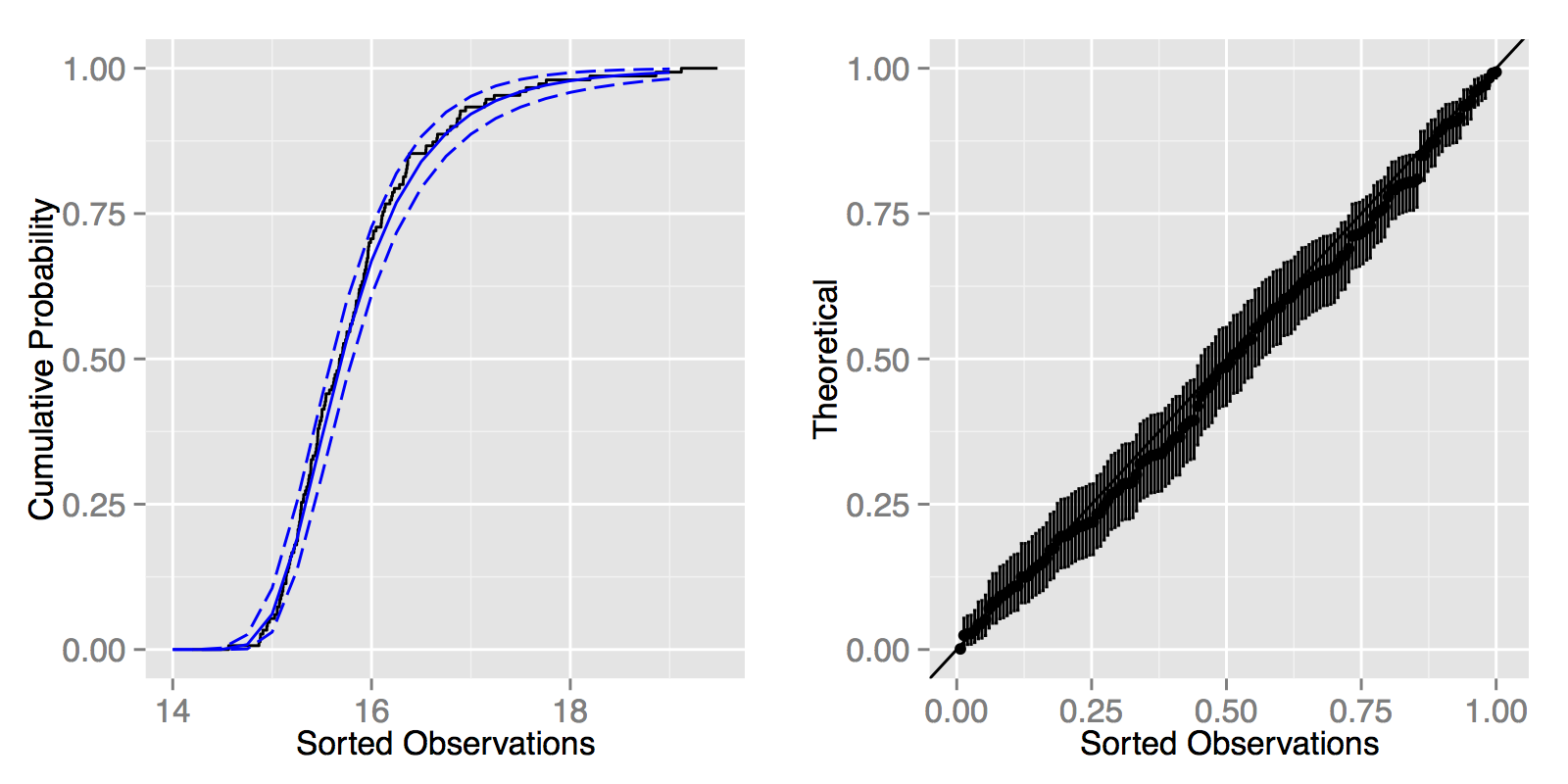}
	\caption{The left panel shows the empirical cumulative distribution  of maximum instantaneous flow from river $j=1$ in January (black solid curve) and the posterior mean of the corresponding posterior cumulative distribution functions (blue solid curve) and corresponding 95\% posterior interval (blue dashed curve). The right panel shows a probability-probability plot of maximum instantaneous flow from river $j=1$ in January, along with 95\% posterior intervals. }
	\label{cdfppp}
\end{figure}

The left panel in Figure \ref{cdfppp} compares the empirical cumulative distribution from river $j=1$ in January with its posterior cumulative distribution functions based on the MCMC runs. The right panel shows the corresponding probability - probability plots. These result indicate that the model describes the data well. Which in turn demonstrates that the MCMC split sampler recaptures the known underlying model, which was used to generate the simulated data. Analogous results hold across all rivers and months (results not shown).

\begin{figure} [hbt]
	\centering
     	\includegraphics[width=1\linewidth]{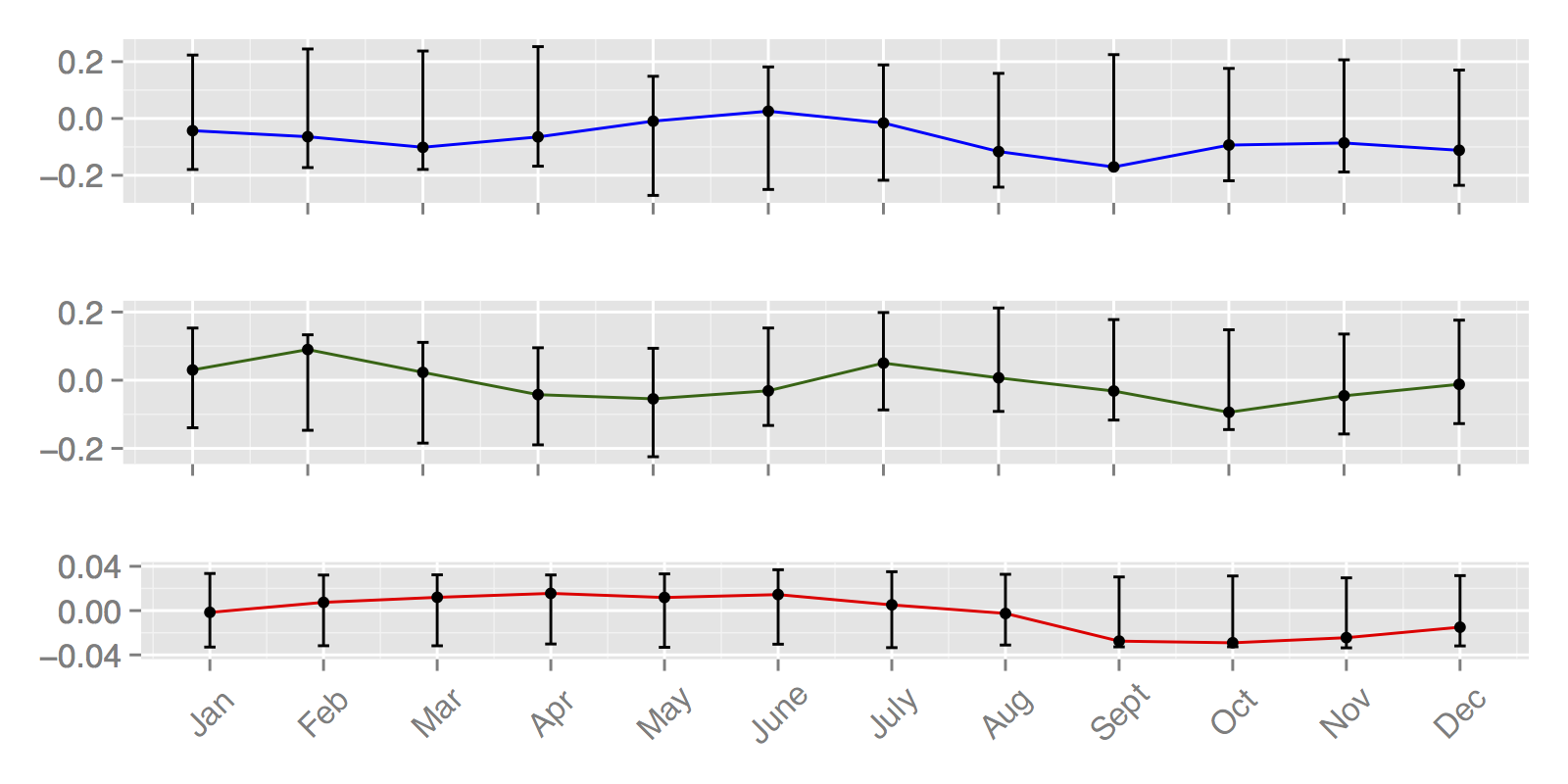}
	\caption{The top panel in shows the known value (denoted with the blue entries) of the seasonal random effect $u_{0,m,\lambda}$ for the log-location parameter $\lambda$ as function of months $m$. The middle panel shows the known value (denoted with the green entries) of the seasonal random effect $u_{0,m,\tau}$ for log-scale parameter $\tau$. The bottom panel shows the known value (denoted with the red entries) of the seasonal random effect $u_{0,m,\xi}$ for the shape parameter $\xi$. The errors bars in all panels represent the corresponding 95\% posterior intervals based on the MCMC-runs. } 
	\label{timeeffect}
\end{figure}

Furthermore, as the data was generated from a known model setup, the results of the inference based on the MCMC runs can be compared to the known values of the model parameters. In Figure \ref{timeeffect}, the known values of the seasonal random effects are shown along with the corresponding 95\% posterior intervals. The top panel in Figure \ref{timeeffect} shows this comparison for the seasonal random effect $u_{0,m,\lambda}$ for the log-location parameter $\lambda$ as a function of months. The middle and the bottom panels in Figure \ref{timeeffect} show the same comparison for $u_{0,m,\tau}$ for the log-scale parameter and $u_{0,m,\xi}$ for the shape parameter $\xi$, respectively. The results reveal that the 95\% posterior intervals for the seasonal random effects contain their known values. These results demonstrate that the MCMC split sampler recaptures the known seasonal random effects. Similar results hold for all other model parameters (results now shown). 

\begin{figure} [hbt]
	\centering
     	\includegraphics[width=0.85\linewidth]{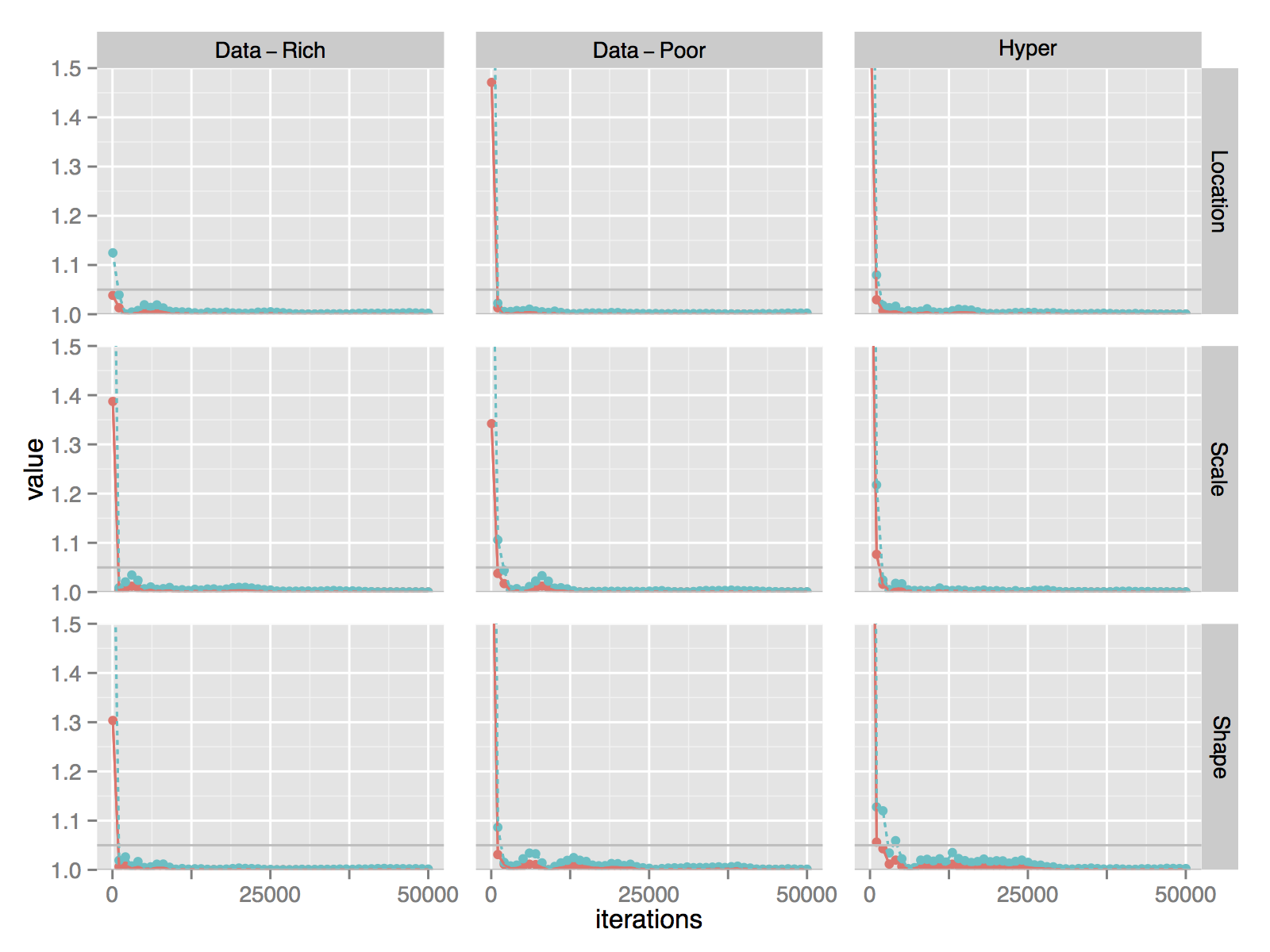}
	\caption{The figure shows Gelman--Rubin plots based on the MCMC run. The first, second and third rows in the first column are based on a randomly chosen log-location parameter $\lambda$; covariate coefficient $\beta_{\lambda}$; and hyperparameter $\psi_{\lambda}$, respectively. The second and third row show an analogous set of parameters based on the log-scale and shape structures, respectively, of the likelihood.}
	\label{GM_2}
\end{figure}

\begin{figure} [hbt]
	\centering
     	\includegraphics[width=0.85\linewidth]{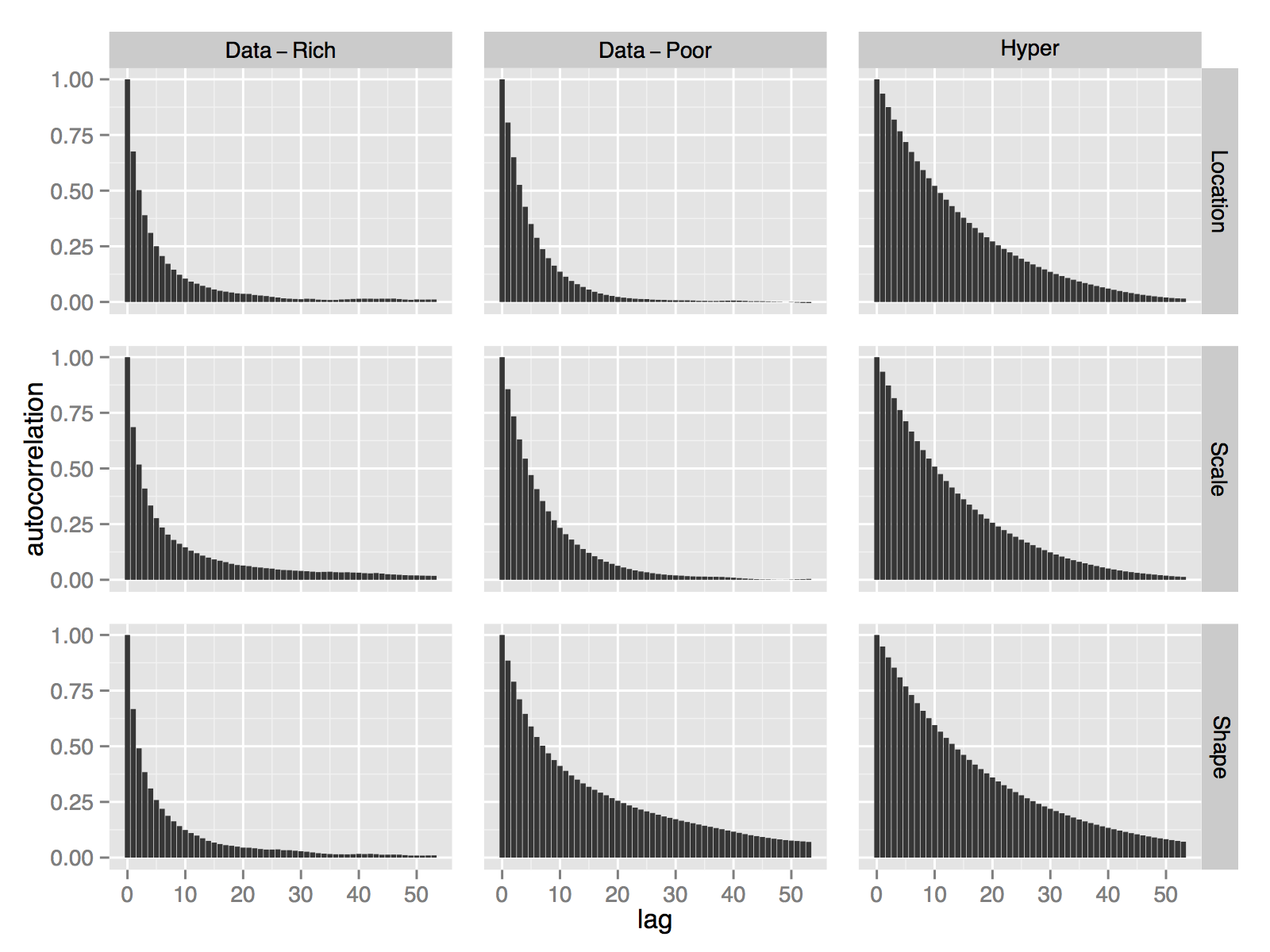}
	\caption{The figure shows auto-correlation plots based on the MCMC run. The first, second and third rows in the first column are based on a randomly chosen log-location parameter $\lambda$; covariate coefficient $\beta_{\lambda}$; and hyperparameter $\psi_{\lambda}$, respectively. The second and third row show an analogous set of parameters based on the log-scale and shape structures, respectively, of the likelihood.}
	\label{Auto_2}
\end{figure}

Gelman--Rubin plots and auto-correlation plots for nine model parameters based on the MCMC run are shown in Figures \ref{GM_2} and \ref{Auto_2}, respectively. Both plots are based on the same set of parameters and arranged identically. Three parameters were chosen from the location, scale and shape structures of the proposed model which were placed in the first, second and third rows of Figures \ref{GM_2} and \ref{Auto_2}, respectively. The first columns are based on parameters from the data-rich part of the latent field; the second columns are based on parameters from the data-poor part of the latent field; and the third column is based on hyperparamters. 

The Gelman--Rubin plots in Figure \ref{GM_2} show that the sampler has converged in the mean after roughly 10.000 iteration. Similar results hold for all the model parameters (results not shown). Furthermore, the autocorrelation plots in Figure \ref{Auto_2} demonstrate that the MCMC chains for the parameters from both the data-rich and the data-poor parts of the latent field, exhibit a negligible autocorrelation after lag 10. The hyperparameters show a negligible autocorrelation after lag 30. Relying on these results, the MCMC chains exhibit all signs of having converged. Moreover, these results further indicate that the MCMC split sampler, with the modified proposal density of \cite{roberts1997weak} implied by equation (\ref{HHH}) for the hyperparameters, is highly computationally efficient in both the data-rich and data-poor blocks. 


%

\section{Discussion}

The main advantage (or novelty) of the MCMC split sampler lies in how the proposed blocking scheme leads to conditional posterior structures within in the two blocks, which can be exploited in order to construct computationally efficient sampling schemes for each block. Additionally, the MCMC split sampler is, in principle, designed as a modular sampling scheme in the sense that any MCMC sampling scheme can be implemented within the blocks. Therefore, in the authors' view, the MCMC split sampler presents an interesting area of future research as new sampling schemes for either block can be developed independently of the other block.  

In the data-rich block, we proposed a Metropolis--Hastings algorithm with an independence proposal density which was constructed with a Gaussian approximation of the conditional posterior density evaluated at its mode.  Furthermore, we proposed a modification the sampler which is applicable if conditional independence assumptions are imposed on the data density function. The modification can potentially increase the computational efficiency of the sampler, as discussed in Appendix \ref{CondInd}. 

Although the proposed sampler in the data-rich block is computationally efficient, it is only applicable in practice if the mode of conditional posterior density function can be found, and can be calculated reasonably fast. For example, in the case of models where each observed data point has more than one unique data density parameters associated with it, say of the type
$
	y_i \sim \pi(y_i | \mu_i, \sigma_i)
$
for every measurement $i$, finding the mode of the conditional posterior $\pi(\mu_i, \sigma_i | y_i )$ becomes computationally impractical in some cases. Models of this type include, for example, certain spatial temporal models \citep{hrafnkelsson2012spatial}.  Similar computational issues also arise if data dependence at the data-level of a LGM is desired, see for example \cite{davison2012statistical} where t-copulas are implemented with g.e.v. marginal density functions at the data level as a model for spatial extremes. However, in both of the aforementioned cases different sampling scheme for the data-rich block can be implemented without changing the sampling scheme of choice in the data-poor block. For example, sampling scheme based on MALA and HMC type algorithm are well suited for the structure of the data-rich block in both cases.


In the data-poor block, the conditional posterior density  $\pi(\fat \nu | \fat \eta, \fat \theta)$ is a Gaussian of the form (\ref{condilater}) and invariant of the data density function. These results serves as one of the main computational advantage introduced by the MCMC split sampler due to the following reasons. First, as the conditional posterior $\pi(\fat \nu | \fat \eta, \fat \theta)$ is Gaussian, a modified version of the one block sampler of \cite{knorr2002block}, which is known to be a highly efficient sampling scheme when applicable \citep{filippone2013comparative}, becomes applicable within the data-poor block regardless of the data-density function at the data level. As a consequence, computationally efficient sampling algorithms can be used to sample from the exact conditional posterior Gaussian density $\pi(\fat \nu | \fat \eta, \fat \theta)$. Further, if the prior density functions in (\ref{priors}) have a sparse GMRF precision structure, then $\pi(\fat \nu | \fat \eta, \fat \theta)$ preserves the sparse structure as discussed in Section \ref{sec:datapoor}, which in turn allows for highly efficient sampling algorithms for the Gaussian density $\pi(\fat \nu | \fat \eta, \fat \theta)$. Therefore, the proposed sampling scheme in Section \ref{sec:datapoor} for the data-poor blocks scales well in terms of computational speed and efficiency with increasing dimensions of the data-poor part of the latent field, which is of great importance to achieve, especially in the field of spatial statistics.

Second, as the conditional posterior density  $\pi(\fat \nu | \fat \eta, \fat \theta)$ is a known Gaussian and the acceptance rate in the sampling scheme for the data-poor block in Section \ref{sec:datapoor} is only dependent on hyperparameters, the computational efficiency of the proposed sampling scheme for the data-poor block is only dependent of the sampling scheme used for the hyperparameters. In this sense, the sampling scheme in Section \ref{sec:datapoor} is in itself modular, that is, any proposal density for the hyperparmeters is applicable. Choosing a computationally efficient sampling scheme for the hyperparameters can thus increase the computational efficiency of the overall sampling scheme within the data-poor block, as demonstrated in the examples in Section \ref{sec:examples}.  That is, in Section \ref{sec:rain} we demonstrated how a proposal density implied by equation (\ref{HavardProp}) may be implemented due to its simplicity. However, in Section \ref{Flood} we proposed a modified version of the sampling scheme of \cite{roberts1997weak} implied by equation (\ref{HHH}), which reduces the autocorrelation in the MCMC chains. In practical terms, the proposal density implied by (\ref{HHH}) can also be implemented Section \ref{sec:rain}, which in turn reduces the autocorrelation in the MCMC chains for the hyperparameters (results omitted).

Due to the modularity of the MCMC split sampler, sampling schemes for the data-rich block can be developed and improved independently of the sampler in the data-poor block, and vice versa. Additionally, as the conditional posterior density $\pi(\fat \nu | \fat \eta, \fat \theta)$ becomes invariant of the data in the data-poor block, the computational advantages introduced by the conditional posterior structure in the data-poor block hold for all LGMs. Moreover, the MCMC split sampler can be applied to various LGMs as it is designed to handle LGMs where latent models Gaussian models are imposed on more than just the mean structure of the data density function. Thus, in our view, further developing and improving sampling schemes that utilize the computational advantages introduced by the MCMC spilt sampler presents an interesting area of future research.

\section*{Acknowledgements}
The authors would like to thank the University of Iceland Doctoral Fund and University of Iceland Research Fund which supported the research. The authors would also like to thank the Icelandic Meteorological Office for providing the data. The authors give their thanks to the Nordic Network on Statistical Approaches to Regional Climate Models for Adaptation (SARMA), especially Prof. Peter Guttorp, for providing travel support. Furthermore, the authors give their thanks to the  Department of Mathematical Sciences at the Norwegian University of Science and Technology for hosting Óli Páll Geirsson several times, and special gratitude to Prof. Håvard Rue for his invitation and valuable conversations.



\bibliographystyle{plain}
\bibliography{bibfile_MCMCSS}

\begin{appendices}
\section{Conditionally independent data density functions} 
\label{CondInd}
In many applications of LGMs, conditional independence assumptions are imposed on the likelihood function, see for example the discussion in \cite{rue2009approximate}. That is, by the model assumptions there exists a partition of $\fat \eta$ into subvectors $\fat \eta_i$, $i=1,\ldots, I$, such that
\begin{align}
\label{ind}
	\pi(\fat y \mid \fat \eta) = \prod_{i=1}^I \pi_i(\fat y_i \mid \fat \eta_i)
\end{align}
where $\pi_i(\fat y_i | \fat \eta_i)$ denotes the marginal data density functions of the $i$-th partition. The conditional independence assumptions in (\ref{ind}) imply $f(\fat \eta) = \sum_i f_i(\fat \eta_{i} )$, where $f_i$ is the logarithm of the marginal data density function $\pi_i(\fat y_i |\fat \eta_i)$.  As discussed in Section \ref{sec:datarich}, the Gaussian approximation in (\ref{GaussApprox}) can be, in some scenarios, a poor approximation of the conditional posterior density in some partition $i$, which in turn may cause the sampler to get stuck and thus lose its efficiency. To address this issue, we suggest a modified version of the sampler proposed in Section \ref{sec:datarich} which retains the computational speed gained by using a Gaussian approximation as a proposal density and is applicable if conditional independence assumptions are imposed on the data density function. Before we introduce the modifications to the sampling scheme in Section \ref{sec:datarich}, we give a few essential technical results for the modifications, which are as follows.

\begin{lemma}
\label{lemma_condind}
Assuming the conditional independence assumptions in (\ref{ind}). Then, every pair of vectors $\fat \eta_{i}$ and $\fat \eta_{i'}$, such that $i\neq i'$, become conditionally independent in the posterior given $(\fat y, \fat \nu, \fat \theta)$. In other words, the following relation holds 
\begin{align}
\label{ohad}
	\pi(\fat \eta_{i} \mid \fat y, \fat \eta_{-i},\fat \nu, \fat \theta) = \pi(\fat \eta_{i} \mid \fat y,\fat \nu, \fat \theta)
\end{align}
for all partitions $i$. 
Furthermore, the conditional posterior density of $\fat \eta_{i}$ is independent of $\fat y_{-i}$, which denotes the subvector of the data vector $\fat y$ that is not observed in partition $i$. That is, the following relationship holds
\begin{align}
\label{augljost2}
	\pi( \fat \eta_i \mid \fat y, \fat \nu, \fat \theta)  = \pi( \fat \eta_i \mid \fat y_i, \fat \nu, \fat \theta) 
\end{align}
for all partitions $i$ 
\end{lemma}
For notational simplicity and based on the relation in (\ref{ohad}) and (\ref{augljost2}), we denote the conditional posterior density  of $\fat \eta_{i}$ with \mbox{$\pi_i(\fat \eta_i \mid \fat y, \fat \nu, \fat \theta)$} henceforth. The following corollary yields a useful identiy for the conditional posterior $\fat \eta_{i}$ with \mbox{$\pi_i(\fat \eta_i \mid \fat y, \fat \nu, \fat \theta)$}.
\begin{corollary}
\label{lemma_cond_structure}
Assume the conditional independence assumptions in (\ref{ind}).  The logaritym of the conditional posterior density \mbox{$\pi_i(\fat \eta_i \mid \fat y, \fat \nu, \fat \theta)$} is 
 \begin{align}
 \label{marginal_cond}
  &\log \pi_i( \fat \eta_i \mid \fat y, \fat \nu, \fat \theta) =  f_i( \fat \eta_i) -\frac{1}{2} \fat \eta_i\trp \fat Q_{\epsilon, (i,i)} \fat \eta_i +\left(\fat Q_{\epsilon, (i,i)} (\fat  Z\fat \nu)_i\right)\trp \fat \eta_i  + K 
\end{align}
where $K$ is a constant. 
\end{corollary}
The subsequent corollary is the immediate  from Theorem \ref{napprox}, Lemma \ref{lemma_condind} and the relation in (\ref{marginal_cond}). 
\begin{corollary}
\label{GaussApproxInd}
Assuming the conditional independence assumptions in (\ref{ind}), the Gaussian approximation of the conditional posterior  \mbox{$\pi_i(\fat \eta_i \mid \fat y, \fat \nu, \fat \theta)$} within each partition $i$ is 
\begin{align}\label{GaussApproxInd_Eq}
  \tilde{\pi}_i(\fat \eta_i \mid \fat y, \fat \theta, \fat\nu) &= \mathcal{N}\left(\fat \eta_i \mid \fat \eta^0_i, \left(\fat Q_{\epsilon} - \fat H \right)_{(i,i)} \right)
\end{align}
where $\fat \eta_i^0$ denotes the mode of the marginal posterior density function $\pi_i(\fat \eta_i \mid \fat y, \fat \nu, \fat \theta)$ and $(\fat Q_{\epsilon} - \fat H )_{(i,i)}$ denotes the submatrix of $(\fat Q_{\epsilon} - \fat H )$ belonging to partition $i$. 
\end{corollary}
The next corollary shows the relation between the Gaussian approximation density functions in (\ref{GaussApprox}) and (\ref{GaussApproxInd_Eq}).
\begin{corollary}
\label{condind_corollary}
Assume the conditional independence assumptions in (\ref{ind}). Furthermore,  let $ \tilde{\pi}(\fat \eta \mid \fat y, \fat \nu, \fat \theta)$ denote the Gaussian approximation of the conditional posterior density \mbox{$\pi(\fat \eta \mid \fat y, \fat \nu,  \fat \theta)$}, and $\tilde{\pi}_i(\fat\eta_i \mid \fat y, \fat\nu, \fat \theta)$ denote the Gaussian approximation of the conditional posterior  \mbox{$\pi_i(\fat \eta_i \mid \fat y, \fat \nu, \fat \theta)$} within partition $i$. Then the following relation holds
\begin{align}
\label{GausRel}
  \tilde{\pi}(\fat \eta \mid \fat y, \fat \nu, \fat \theta) = \prod_i^I \tilde{\pi}_i(\fat\eta_i \mid \fat y, \fat\nu, \fat \theta).
\end{align}
\end{corollary}

The modifications to the sampler proposed in Section \ref{sec:datarich} are based on the following observations. From the conditional posterior independence relation in (\ref{ohad}) in Lemma \ref{lemma_condind} follows directly
\[
	\pi(\fat \eta_{i} \mid \fat y, \fat \eta_{-i},\fat \nu, \fat \theta) = \pi(\fat \eta_{i} \mid \fat y,\fat \nu, \fat \theta)
\]
for all partitions  $i$. It is therefore equivalent to update $\fat \eta_{i} | \fat y, \fat \eta_{-i},\fat \nu, \fat \theta$  iteratively over partitions $i$, with a Gibbs sampling approach using (\ref{GaussApproxInd_Eq})  as a proposal density and to update $\fat \eta_{i} | \fat y,\fat \nu, \fat \theta$ separately over partitions $i$. By updating separately as opposed to iteratively, the number of functions calls is reduced, which in turn reduces computational cost. 

However, in practical terms it is faster to compute the mode once by computing the maximum of $\log \pi(\fat \eta | \fat y, \fat \nu, \fat \theta)$ than computing the maximum of $\log \pi_i(\fat \eta_i | \fat y, \fat \nu, \fat \theta)$ separately  in every partition $i$. This is due to the fact that number of function calls increases as $I$ increases in the numerical optimizing methods when finding the mode of $\log \pi_i(\fat \eta_i | \fat y, \fat \nu, \fat \theta) $ within every partition $i$. Furthermore, the computational cost of calculating the function $ \log \pi ( \fat \eta \mid \fat y, \fat \nu, \fat \theta)$  in (\ref{logdens1}) is minimal and scales well as the dimension of the data-poor block increases, as the $\fat Q_{\epsilon}$ is a diagonal matrix and $\fat Z$ is a fixed sparse matrix. Thus, calculating the gradient and the Hessian matrix of the conditional posterior is also computationally fesiable in many cases.

The relation in (\ref{GausRel}) demonstrates that it is equivalent to propose a new vector $\fat \eta^*$ from the normal approximation $\tilde{\pi}(\fat \eta \mid \fat y, \fat \nu, \fat \theta)$ and to propose new vectors $\fat \eta^*_i$ separately from $ \tilde{\pi}_i(\fat\eta_i \mid \fat y, \fat\nu, \fat \theta)  $ for every $i$. Therefore, to reduce computational cost and to increase computational efficiency, we propose the following modifications the the sampling scheme	in Section \ref{sec:datarich}.  That is, propose a new vector  $\fat \eta^{*}$ from $q(\fat \eta) = \tilde{\pi}(\fat \eta \mid \fat y, \fat \nu, \fat \theta) $ and accept  and reject $\fat \eta^{*}_i$  within each partition separately with the probability 
\begin{align}\label{ra2}
  \alpha_i = \min\left\{1,\frac{ \pi ( \fat\eta_i^* \mid \fat y, \fat \nu, \fat \theta)}{\tilde{\pi}(\fat\eta_i ^ * \mid \fat y, \fat \nu, \fat \theta) } \bigg / 
  \frac{ \pi ( \fat\eta_i^k \mid \fat y, \fat \nu, \fat \theta) }{\tilde{\pi}(\fat\eta_i ^ k \mid \fat y, \fat \nu, \fat \theta)} \right\}.
\end{align}
The acceptance ratio in (\ref{ra2}) can be calculated separately over partitions $i$ with a low computational cost, as demonstrated in the following corollary.
\begin{corollary}
\label{opgcor}
Assume the conditional independence assumptions in (\ref{ind}) and adopt the same notation as in Lemma \ref{lemmaratio1}. The logarithm of the acceptance ratio in (\ref{ra2}), that is
\begin{align*}
  r_i = \log \left(\frac{ \pi ( \fat\eta_i^* \mid \fat y, \fat \nu, \fat \theta)}{\tilde{\pi}(\fat\eta_i ^ * \mid \fat y, \fat \nu, \fat \theta) } \bigg / 
  \frac{ \pi ( \fat\eta_i^k \mid \fat y, \fat \nu, \fat \theta) }{\tilde{\pi}(\fat\eta_i ^ k \mid \fat y, \fat \nu, \fat \theta)}\right)
\end{align*}
where $\tilde{\pi}_i(\fat\eta_i \mid \fat y, \fat\nu, \fat \theta)$ denotes the Gaussian approximation of the conditional posterior  \mbox{$\pi_i(\fat \eta_i \mid \fat y, \fat \nu, \fat \theta)$} within partition $i$. Then $r_i$ can be simplified to
 \[
  r_i =f_i( \fat \eta_i^*) + \fat \rho (\fat \eta^*)_i\trp \fat 1 - (f_i( \fat \eta_i^k) + \fat \rho (\fat \eta^k)_i\trp \fat 1 )
\]
for all $i$, where $f_i$ is the logarithm of the marginal data density function in partition $i$ and
\[
  \fat \rho(\fat \eta) = \left(\frac1 2 \fat \eta\trp \fat H   + \fat b\trp \right) \circ \fat \eta
\]
for notational simplicity, where $\circ$ denotes an entrywise multiplication. The corresponding acceptance probability in partition $i$ is thus
\begin{equation*}
  \alpha_i = \min\left\{ 1, \exp r_i \right\}.
\end{equation*}
\end{corollary}
The sampling scheme for the data-rich block with the aforementioned modifications is summarised in Algorithm 2


\section{Proofs}
\label{Proofs}
\subsection{Proof of Lemma \ref{setnA}}
\label{Proof_Gauss}
\begin{proof}
By know results about Gaussian distributions and inverses of block matrices, the joint distribution of $(\fat \eta, \fat \nu)$ is given by
\begin{align*}
  \pi{\Matrix{\fat \eta \\ \fat \nu}}  &=\ndist{{\Matrix{\fat \eta \\ \fat \nu}} \bigg|  \Matrix{\fat Z\fat{\mu}_\nu\\ \fat{\mu}_\nu}}{\Matrix{\fat{Q}_\epsilon & -\fat{Q}_\epsilon\fat Z \\ -\fat Z\trp\fat{Q}_\epsilon & \fat{Q}_\nu+\fat Z\trp\fat{Q}_\epsilon\fat Z}\inv} \notag \\ 
  &=\ndist{{\Matrix{\fat \eta \\ \fat \nu}} \bigg|  \Matrix{\fat Z\fat{\mu}_\nu\\ \fat{\mu}_\nu}}{\Matrix{\fat{Q}_\epsilon\inv + \fat Z \fat Q_\nu\inv \fat Z\trp &  \fat Z \fat Q_\nu\inv \\  \fat Q_\nu\inv \fat Z\trp & \fat Q_\nu\inv}}
\end{align*}
The conditional distribution  $\fat \nu$ conditioned on $\fat \eta $ follows directly from Lemma 2.1 in \citep{rue2005gaussian}, that is, 
\begin{align*}
  &\pi(\fat \nu \mid \fat \eta)  =  \mathcal N  \left( \fat\nu \Big | \fat Q_{\nu|\eta} ^{-1}(\fat Q_\nu \fat \mu_\nu + \fat Z\trp \fat Q_\epsilon \fat\eta) , \fat Q_{\nu|\eta} ^{-1}\right)
\end{align*}
where $\fat Q_{\nu|\eta} = \fat{Q}_\nu+\fat Z\trp\fat{Q}_\epsilon\fat Z$.
\end{proof}

\subsection{Proof of Theorem \ref{napprox}}
\label{Proof_lemma_datarich}
\begin{proof}
The conditional posterior density function $\pi(\fat \eta \mid \fat y, \fat \nu,  \fat \theta) $ is proportional to the product of the data density function and the conditional Gaussian prior density $\pi(\fat \eta \mid \fat \nu, \fat \theta)$  given by (\ref{priors}), that is 
\begin{align*}
\pi(\fat \eta \mid \fat y, \fat \nu,  \fat \theta) \propto \pi(\fat y \mid \fat \eta)\pi(\fat \eta \mid \fat \nu, \fat \theta).
\end{align*} 
Thus, the logarithm of the conditional posterior density function is given by
 \begin{align*}
  \log \pi ( \fat \eta \mid \fat y, \fat \nu, \fat \theta) =f( \fat \eta) -\frac{1}{2} \fat \eta\trp \fat Q_{\epsilon} \fat \eta +(\fat Q_{\epsilon} \fat Z\fat \nu)\trp \fat \eta  + \text{const}
\end{align*}
where $f( \fat \eta) = \log \pi (\fat y\mid \fat \eta)$ for notational convenience. The second order Taylor approximation of $f( \fat \eta)$ expanded around the mode $\fat \eta^0$ of the conditional posterior $\pi( \fat \eta \mid \fat y, \fat \nu, \fat \theta)$ is
\begin{align*}
	f( \fat \eta) &\approx f( \fat \eta^0) + \nabla f(\fat \eta^0)\trp(\fat \eta - \fat\eta^0) 
	+ \frac 12 (\fat \eta - \fat\eta^0)\trp \fat H(\fat \eta - \fat\eta^0) \\
	&= \frac 12 \fat\eta\trp \fat H \fat\eta + (\nabla f(\fat \eta^0) - \fat H \fat \eta^0)\trp \fat \eta + \text{const}.
\end{align*}
Consequently, the second order Taylor approximation of  $\log \pi ( \fat \eta \mid \fat y, \fat \nu, \fat \theta) $ expanded around $\fat \eta^0$ becomes
 \begin{align*}
  \log \pi ( \fat \eta \mid \fat y, \fat \nu, \fat \theta)   &\approx\frac 12 \fat\eta\trp \fat H \fat\eta + (\nabla f(\fat \eta^0) - \fat H \fat \eta^0)\trp \fat \eta -\frac{1}{2} \fat \eta\trp \fat Q_{\epsilon} \fat \eta +(\fat Q_{\epsilon} \fat Z\fat \nu)\trp \fat \eta  + \text{const}  \\ 
  & =-\frac 1 2 \fat \eta\trp \left( \fat Q_{\epsilon} -\fat H\right)\fat \eta + (\fat Q_{\epsilon} \fat Z\fat \nu+ \fat b)\trp \fat \eta + \text{const},
\end{align*}
where $\fat b = (\nabla f(\fat \eta^0) - \fat H \fat \eta^0)$. This derivation yields a Gaussian approximation with a mean vector 
\[
\left( \fat Q_{\epsilon} -\fat H\right)^{-1}(\fat Q_{\epsilon} \fat Z\fat \nu+ \fat b)
\] 
and covariance matrix $\left( \fat Q_{\epsilon} -\fat H\right)\inv$. However, as the vector $\fat \eta^0$ is the mode of the conditional posterior function $\pi( \fat \eta \mid \fat y, \fat \nu, \fat \theta)$ the following relation holds
\begin{align*}
	\nabla  \log \pi &( \fat \eta^0 \mid \fat y, \fat \nu, \fat \theta) = \nabla f(\fat\eta^0) - \fat Q_\epsilon \fat \eta_0 + (\fat Q_{\epsilon} \fat Z\fat \nu)\trp  = \fat 0.
\end{align*}
The mean of the Gaussian approximations becomes
\begin{align*}
	&\left( \fat Q_{\epsilon} -\fat H\right)^{-1}(\fat Q_{\epsilon} \fat Z\fat \nu+ \fat b)\\
	&=\left( \fat Q_{\epsilon} -\fat H\right)^{-1}(\fat Q_{\epsilon} \fat Z\fat \nu+ \nabla f(\fat \eta^0) - \fat H \fat \eta^0 ) \\
	&=\left( \fat Q_{\epsilon} -\fat H\right)^{-1}(\fat Q_{\epsilon} \fat\eta^0 - \fat H \fat \eta^0 ) \\
	&=\left( \fat Q_{\epsilon} -\fat H\right)^{-1}(\fat Q_{\epsilon} - \fat H \fat )\fat \eta^0 
	= \fat \eta^0
\end{align*} 
Thus, a Gaussian approximation of the conditional posterior density function $\pi( \fat \eta \mid \fat y, \fat \nu, \fat \theta)$ evaluated at the mode $\fat \eta^0$ is given by
\begin{align*}
\tilde{\pi}(\fat \eta \mid \fat y, \fat \nu, \fat \theta) &= \mathcal{N} \left(\fat \eta \mid \fat\eta^0, (\fat Q_{\epsilon} -\fat H)\inv \right).
\end{align*}
\end{proof}

\subsection{Proof of Lemma \ref{lemmaratio1}}
\begin{proof}
The logarithm of the acceptance ratio given in (\ref{ra1}) is
\begin{align}
\label{ra1A}
  r= \log \frac{ \pi ( \fat \eta^* \mid \fat y, \fat \nu, \fat \theta)q(\fat \eta ^ k)}{ \pi ( \fat \eta^k \mid \fat y, \fat \nu, \fat \theta) q(\fat \eta ^ *) }
\end{align} 
where $\pi ( \fat \eta \mid \fat y, \fat \nu, \fat \theta)$ is the conditional posterior density function given in (\ref{priors}) and  $q(\fat \eta )$ is the proposal density based on the Gaussian approximation in (\ref{GaussApprox}). The right hand side term in (\ref{ra1A}) can be written as
\begin{align*}
	\log \pi ( &\fat \eta^* \mid \fat y, \fat \nu, \fat \theta) - \log q(\fat \eta^*)  \\
	&-\left(\log \pi ( \fat \eta^k \mid \fat y, \fat \nu, \fat \theta) + \log   q(\fat \eta^k)\right)
\end{align*}
Since the proposal density $q$ is based on the Gaussian approximation in (\ref{GaussApprox})  the following holds
\begin{align*}
	\log \pi ( \fat \eta \mid \fat y, \fat \nu, \fat \theta) - \log q(\fat \eta) 
	&=f( \fat \eta) -\frac{1}{2} \fat \eta\trp \fat Q_{\epsilon} \fat \eta +(\fat Q_{\epsilon} \fat Z\fat \nu)\trp \fat \eta\\
	 &\quad -\left( \frac 12 \fat\eta\trp \fat H \fat\eta + \fat b\trp \fat \eta -\frac{1}{2} \fat \eta\trp \fat Q_{\epsilon} \fat \eta +(\fat Q_{\epsilon} \fat Z\fat \nu)\trp \fat \eta   \right) + \text{const} \\
	&=f( \fat \eta)  - \left( \frac 12 \fat\eta\trp \fat H \fat\eta + \fat b\trp \fat\eta\right) + \text{const}
\end{align*}
which yields the results in (\ref{r1}).
\end{proof}


\subsection{Proof of Lemma \ref{maindatapoor}}
\begin{proof}
By definition of the proposal density in (\ref{sampla1}) the following holds
\begin{align*}
\frac{q(\fat \nu^k, \fat \theta^k\mid\fat \nu^*, \fat \theta^*)}{q(\fat \nu^*, \fat \theta^*\mid\fat \nu^k, \fat \theta^k)}
  =\frac{\pi (\fat \nu^k \mid \fat \eta^{k+1},\fat \theta^k)q(\fat \theta^k \mid \fat \theta^*)}{\pi (\fat \nu^* \mid \fat \eta^{k+1}, \fat \theta^*)q(\fat \theta^* \mid \fat \theta^k)},
\end{align*}
where $q(\fat \theta^* | \fat \theta^k)$ is some proposal density for $\fat \theta$ and $\pi (\fat \nu \mid \fat \eta, \fat \theta)$ is the conditional Gaussian density function in (\ref{condilater}) in Lemma \ref{setnA}. Therefore, the acceptance ratio in (\ref{allratio}) can be written as
\begin{align}
\label{r4}
&\frac{\pi( \fat \nu^*, \fat \theta^*\mid \fat y, \fat \eta^{k+1}) }{\pi( \fat \nu^k, \fat \theta^k\mid \fat y, \fat \eta^{k+1})}
\frac{\pi (\fat \nu^k \mid \fat \eta^{k+1},\fat \theta^k)}{\pi (\fat \nu^* \mid \fat \eta^{k+1}, \fat \theta^*)}\frac{q(\fat \theta^k \mid \fat \theta^*)}{q(\fat \theta^* \mid \fat \theta^k)}=\frac{\pi(\fat \theta^* \mid\fat\eta^{k+1}) }{\pi(\fat \theta^k \mid\fat\eta^{k+1}) }
   \frac{q(\fat \theta^k \mid \fat \theta^*)}{q(\fat \theta^* \mid \fat \theta^k)} 
\end{align}
since $\pi(\fat \nu, \fat \theta | \fat y, \fat\eta)$ $= \pi(\fat \nu, \fat \theta |\fat\eta)$, as discussed in Section \ref{sec:datapoor}, and  ${\pi(\fat \nu, \fat \theta \mid \fat \eta)}/{\pi(\fat \nu \mid \fat \eta, \fat \theta)}= \pi(\fat\theta\mid\fat \eta)$ for any $\fat \eta, \fat \nu$ and $\fat \theta$.
The result in (\ref{r4}) demonstrates that the acceptance ratio in (\ref{allratio}) is only dependant on $\fat \theta$ within in the proposed setup. In other words, the acceptance ratio in (\ref{allratio}) becomes independent of the value of $\fat \nu$. 
\end{proof}

\subsection{Proof of Theorem \ref{maindatapoor2}}
\begin{proof}
In order to rewrite  $\pi(\fat \theta \mid\fat\eta )$ in (\ref{testa}) we use the relation 
\begin{align}
\label{rat1}
  \pi(\fat \theta \mid\fat\eta ) \propto \pi(\fat \theta) \pi(\fat \eta \mid \fat \theta).
\end{align}
Further, by the law of conditional probability, the following holds
\begin{align}
\label{rat2}
  \pi(\fat \eta \mid \fat \theta) = \frac{\pi(\fat \eta, \fat \nu \mid \fat \theta)}{\pi(\fat\nu \mid \fat \eta, \fat \theta)} = \frac{\pi( \fat \eta \mid \fat \nu, \fat \theta) \pi( \fat \nu \mid \fat \theta) }{\pi( \fat \nu \mid\fat \eta, \fat \theta) }
\end{align}
As $\pi(\fat \eta \mid \fat \theta) $ is independent of the value of $\fat \nu$, it follows that the two ratios in (\ref{rat2}) are invariant of the choice of $\fat \nu$. In particular, the following holds
\begin{align}
\label{ratiowith0}
  \pi(\fat \eta \mid \fat \theta)  = \frac{\pi( \fat \eta \mid \fat 0, \fat \theta) \pi( \fat 0 \mid \fat \theta) }{\pi( \fat 0 \mid\fat \eta, \fat \theta) }
\end{align}
by choosing the value $\fat \nu = \fat 0$. Combining (\ref{rat1}) and (\ref{ratiowith0}) yields 
\begin{align*}
&\frac{\pi(\fat \theta^* \mid\fat\eta^{k+1}) }{\pi(\fat \theta^k \mid\fat\eta^{k+1}) } = \frac{\pi(\fat \theta^*) \pi(\fat \eta^{k+1} \mid \fat \theta^*) }{\pi(\fat \theta^k) \pi(\fat \eta^{k+1} \mid \fat \theta^k) }\nonumber\\
  &=\frac{\pi (\fat \theta ^*)}{\pi(\fat \theta ^k)}\times
  \frac{\pi( \fat \eta^{k+1} \mid \fat 0, \fat \theta^*) \pi( \fat 0 \mid \fat \theta^*) }{\pi( \fat 0 \mid\fat \eta^{k+1}, \fat \theta^*) } \nonumber\times\frac{\pi( \fat 0 \mid\fat \eta^{k+1}, \fat \theta^k) }{\pi( \fat \eta^{k+1} \mid \fat 0, \fat \theta^k) \pi( \fat 0 \mid \fat \theta^k) }.
\end{align*} 
Moreover, if the Gaussian prior density functions in  (\ref{priors}) are GMRFs with sparse precision structures, then all of the conditional density functions on the rightmost side of (\ref{rat2}) are GMRFs with sparse precision structures, by known results about conditioning on subvectors as demonstrated in Theorem 2.5 in \cite{rue2005gaussian}. 


\end{proof}

\subsection{Proof of Lemma \ref{lemma_condind}}
\begin{proof}
As the matrix $\fat Q_{\epsilon}$ is a diagonal matrix the following holds
\begin{align}
\label{NormMult10}
	\pi(\fat \eta \mid \fat \nu, \fat \theta) &= \mathcal N \left(\fat \eta \big| \fat Z \fat\nu, \fat Q_{\epsilon} \inv  \right) \nonumber \\
	&=\prod_{i=1}^I \mathcal N \left(\fat \eta_{i}\big| (\fat Z \fat\nu)_i , \fat Q_{\epsilon, (i,i)} \inv  \right) 
\end{align}
where $\fat Q_{\epsilon, (i,i)}$ denotes the submatrix of $\fat Q_{\epsilon}$ belonging to partition $i$.  Let 
\[
	\pi_i (\fat \eta_i | \fat \nu, \fat \theta) = \mathcal N \left(\fat \eta_{i}\big| (\fat Z \fat\nu)_i , \fat Q_{\epsilon, (i,i)} \inv  \right) 
\]
which serves as the conditional prior density function for the data-rich part of the latent field belonging to partition $i$.  The relation in (\ref{NormMult10}) along with the conditional independence assumptions in (\ref{ind}) yield
\begin{align*}
\pi(\fat \eta \mid \fat y,  \fat \nu,  \fat \theta) &\propto \pi(\fat y \mid \fat \eta) \pi(\fat \eta \mid \fat \nu, \fat \theta) \nonumber \\
&= \prod_{i=1}^I \pi_i(\fat y_i \mid \fat \eta_i) \pi_i (\fat \eta_i | \fat \nu, \fat \theta) 
\end{align*} 
which demonstrates that the vectors $\fat \eta_{i}$ and $\fat \eta_{i'}$ are conditionally independent in the conditional posterior given $(\fat y, \fat \nu, \fat \theta)$, for all $i \neq i'$.  

Furthermore, the conditional independence assumptions in (\ref{ind}) also yield
\begin{align}
\label{augljost20}
\pi(\fat \eta_i \mid \fat y,  \fat \nu,  \fat \theta)  \nonumber 
&\propto \pi(\fat y_i \mid \fat \eta_i)\pi_i(\fat \eta_{i} \mid \fat \nu, \fat \theta) \\
&\propto \pi(\fat \eta_i \mid \fat y_i,  \fat \nu,  \fat \theta)
\end{align} 
which demonstrates that $\pi(\fat \eta_i \mid \fat y,  \fat \nu,  \fat \theta)$ is independent of $\fat y_{-i}$.  
\end{proof}

\subsection{Proof of Corollary \ref{lemma_cond_structure}}
\begin{proof}
The conditional independence assumptions in (\ref{ind}), the relation in (\ref{NormMult10}) and the relation in (\ref{augljost20}) yield
 \begin{align*}
  \log \pi_i( \fat \eta_i \mid \fat y_i, \fat \nu, \fat \theta)  
  &=\log \pi(\fat y_i \mid \fat \eta_i) + \log \pi_i(\fat \eta_{i} \mid \fat \nu, \fat \theta) + K\\
  &=f_i( \fat \eta_i) -\frac{1}{2} \fat \eta_i\trp \fat Q_{\epsilon, (i,i)} \fat \eta_i +\left(\fat Q_{\epsilon, (i,i)} (\fat  Z\fat \nu)_i\right)\trp \fat \eta_i  + K
\end{align*}
for every partition $i$. 
\end{proof}

\subsection{Proof of Corollary \ref{GaussApproxInd}}
\begin{proof}
The result follows by using Theorem \ref{napprox} on the $\log \pi_i( \fat \eta_i \mid \fat y, \fat \nu, \fat \theta)$ given in  (\ref{marginal_cond}) instead of  $\log \pi( \fat \eta | \fat y, \fat \nu, \fat \theta)$.
\end{proof}

\subsection{Proof of Corollary \ref{condind_corollary}}
\begin{proof}
The relation in (\ref{ohad}) and (\ref{augljost2}) in Lemma (\ref{lemma_condind})  implies that the elements of mode $\fat \eta^0$ belonging to partition $i$ are also the mode of $\pi_i(\fat \eta_i | \fat y_i, \fat \nu, \fat \theta)$ in every partition $i$. The results then follows from Lemma \ref{lemma_condind}, Theorem \ref{napprox}  and Corollary  \ref{GaussApproxInd}.
\end{proof}

\subsection{Proof of Corollary \ref{opgcor}}
\begin{proof}
As conditional independence are imposed over partitions $i$, Lemma \ref{lemma_condind} yields the following

\begin{align*}
  \log\pi ( \fat \eta \mid \fat y, \fat \theta, \fat \nu) - \log\tilde{\pi}(\fat \eta \mid \fat y, \fat \nu, \fat \theta) 
  &=  f( \fat \eta) -  \frac1 2 \fat \eta\trp \fat H \fat \eta  - \fat b\trp \fat \eta  + \text{const}\\ \notag
  &= \sum_i \left(f_i( \fat \eta_i) -  \frac{1}{2}\fat \eta_i\trp \fat H_{(i,i)} \fat \eta_i  - \fat b_i\trp \fat \eta_i\right)+ \text{const}.
\end{align*}
The results follows by similar derivations as in the proof of Lemma \ref{lemmaratio1}.
\end{proof}

\end{appendices}

\end{document}